\documentclass[a4paper,11pt]{article}

\usepackage{jheppub} 

\usepackage[T1]{fontenc} 
\usepackage{amssymb}
\usepackage{amsthm}
\usepackage{amsopn}
\usepackage{amsmath}
\usepackage{graphicx}
\usepackage{color}
\usepackage{float}
\usepackage{subfigure}
\usepackage{multirow}
\usepackage{hyperref}
\usepackage[normalem]{ulem} 
\usepackage{soul} 

\DeclareMathOperator{\sech}{\mathrm{sech}}

\title{Kink Dynamics in a Parametric $\phi^6$ System: A Model With Controllably Many Internal Modes}

\author[a,1]{A. Demirkaya,\note{Corresponding author.}}
\author[a]{R. Decker,}
\author[b]{P. G. Kevrekidis}
\author[c,d]{I. C. Christov}
\author[c]{A. Saxena}

\affiliation[a]{Mathematics Department, University of Hartford,\\200 Bloomfield Ave, West Hartford, CT 06117, USA}
\affiliation[b]{Department of Mathematics and Statistics, University of Massachusetts,\\Amherst, MA 01003-4515, USA}
\affiliation[c]{Center for Nonlinear Studies and Theoretical Division, Los Alamos
National Laboratory,\\Los Alamos, NM 87545, USA}
\affiliation[d]{School of Mechanical Engineering, Purdue University,\\ West Lafayette, IN 47907, USA}

\emailAdd{demirkaya@hartford.edu}
\emailAdd{rdecker@hartford.edu}
\emailAdd{kevrekid@math.umass.edu}
\emailAdd{christov@purdue.edu}
\emailAdd{avadh@lanl.gov}

\abstract{We explore a variant of the $\phi^6$ 
model originally proposed in Phys.\ Rev.\ D {\bf 12}, 1606 (1975) as
a prototypical, so-called, ``bag'' model in which domain walls play the
role of quarks within hadrons. We examine the 
steady state of the model, namely an apparent bound state of
two kink structures. We explore its linearization, and we find that,
as a function of a parameter controlling the curvature
of the potential, an {\it effectively arbitrary} number of internal
modes may arise in the point spectrum of the linearization about the
domain wall profile.
We explore some of the key characteristics of kink-antikink
collisions, such as the critical velocity and the multi-bounce
windows, and how they depend on the principal parameter
of the model. We find that the critical velocity exhibits
a non-monotonic dependence on the parameter controlling the curvature
of the potential. For the multi-bounce windows, we find that their range and
complexity decrease as the relevant parameter decreases
(and as the number of internal modes in the model increases).
We use a modified collective coordinates method [in the spirit
of recent works such as {Phys.\ Rev.\ D} {\bf 94}, 085008 (2016)] 
in order to capture the relevant phenomenology in a semi-analytical
manner.}

\begin{document} 
\maketitle
\flushbottom

\section{Introduction}
\label{sec:intro}

The study of field theories of the nonlinear Klein--Gordon type and especially
of the general class of $\phi^4$ models is a topic of wide
appeal and time-honored history~\cite{belova}. Such models are of interest to a broad array of applications.
These range from describing domain walls in cosmology \cite{anninos,vilenkin01} to structural phase transitions \cite{gufan,behera}, uniaxial
ferroelectrics or even simple polymeric chains \cite[Ch.~9]{Vach}; see also 
Refs.~\cite{manton,campbell} and those therein. Note that the usual $\phi^6$ model is invoked in the description of first-order transitions \cite{gufan} 
in ferroelectric \cite{ferroelectric}, ferroelastic and magnetoelastic \cite{planes} crystals, the nematic-to-isotropic transition in liquid crystals \cite{liquid crystal}, the electroweak 
transition in the early Universe \cite{universe} and related field theoretic contexts \cite{makhankov}. A particularly 
appealing feature that was discovered early on
was the existence of a fractal structure \cite{anninos} in
the collisions between the fundamental nonlinear wave structures
(a kink and an antikink) in such models. This is a topic that was
initiated by the numerical investigations in \cite{campbell,Campbell1} (see also \cite{belova}), and it
is still under active investigation both in the physics community (see, e.g., 
Refs.~\cite{gani1,simas,danial} and those therein) and the mathematics
community (see, e.g., the mathematical analysis of the relevant mechanism 
provided in \cite{goodman2}).

On the physics side, there has been a very extensive array
of recent studies of different classes of phenomena that challenge
many of the traditional perspectives of this problem. The
more standard view has been that the internal mode of
the $\phi^4$ kink and its resonant dynamics with other
(e.g., translational and extended) degrees of freedom
are responsible for the observation of the multi-bounce
windows~\cite{anninos,campbell,Campbell1}. However, in
recent studies all sorts of ``anomalies'' have arisen.
For instance, models of the $\phi^6$ type have been
shown to feature multi-bounce windows in the absence
of internal modes~\cite{shnir1}. Parametric deformations of
the $\phi^4$ model that introduce additional internal
modes have been claimed to suppress two-bounce windows
(in which the kinks do not separate upon colliding
once, but twice)~\cite{simas}. Another perhaps even
more troubling feature has stemmed from the work
presented in Refs.~\cite{weigel,weigel2}. These claim that if all terms are included in the collective coordinate models
aiming at an effective description of
these collisions and originating from (for example) the work of Sugiyama \cite{sugiyama}
(see also~\cite{anninos}), then
significant problems (both practically regarding the computation
and, more importantly, regarding the results and conclusion drawn thereof) emerge when attempting a quantitative comparison with direct numerical simulations of the governing partial differential equation (PDE).
In addition to all these studies, the broadening phenomenology of kink interactions has also enhanced the interest in studying models of the $\phi^6$,
$\phi^8$, $\phi^{10}$ and even $\phi^{12}$
types~\cite{gani1,gani2,khare,gani3}. A well-rounded, recent summary
encompassing a large volume
of works on this theme can be found in~\cite{danial}.
As an aside, it is worthwhile to note in passing
  that fluctuations (considered also below) in the case of the
  $\phi^6$ model have been recently examined not only at the
  classical level but also at the quantum level.
  { For instance, the works of~\cite{alonso} and~\cite{weig}
  proposed two different procedures for evaluating the quantum
  corrections to the kink mass.}

The aim of the present work is to examine a different type of $\phi^6$ model than, e.g., the one 
studied in most works on this topic (Refs.~\cite{shnir1,gani1,danial,lohe} and those therein).
On the one hand, the model we consider is a variant of the usual $\phi^6$ model that is potentially
of intrinsic interest in its own right within high-energy physics, as it was
originally proposed by Christ and Lee~\cite{lee1}, where it was presented as
a prototypical, so-called, ``bag'' model in which domain walls play the
role of quarks within hadrons.  Christ and Lee~\cite{lee1} quantized the one-dimensional 
scalar field theory whose classical solutions include soliton pairs. This variant has a significant ``advantage'' over more ``rigid'' (non-parametric)
forms of the model employed in the works of~\cite{lohe,shnir1,gani1,danial},
in that it possesses a tunable parameter $\epsilon$, and a topological solitary wave (kink)
exists for {\it all} values of this parameter.
Even more importantly for our purposes, as we will see below,
the solitary wave
contains a {\it controllable number} of internal modes as the parameter $\epsilon$
is varied, and this number progressively grows as 
$\epsilon \rightarrow 0$. In that light, the $\phi^6$ variant from \cite{lee1} is
an excellent platform for exploring the role of internal modes in
kink-antikink collision dynamics, the variation of the critical
velocity, the formation of multi-bounce windows and all the related
notions. Thus, it is exactly this effort that we undertake in this work.

This paper is organized as follows. In section~\ref{sec:model},
we discuss the general background of the model, following~\cite{lee1}
and providing the details of the single kink and its excitation
spectrum. Then, in section~\ref{sec:NR}, we explore numerically 
the collisional dynamics of kinks and antikinks, for different values of the model's parameter $\epsilon$, as well as the dependence of the
critical velocity (for the ultimate separation of the kink and antikink) as a function of
the parameter $\epsilon$. In section~\ref{sec:CC}, we
connect these purely numerical results to a collective coordinate (CC) description of the phenomenon.
In applying the CC theory in its ``standard'' form, as stemming
from the early work of~\cite{sugiyama}, we encountered
the same types of problems that have been recently reported
in~\cite{weigel,weigel2}. Therefore, we have opted to make the types
of amendments/modifications as in the CC approach proposed in \cite{weigel,weigel2}. Most notably,
these involve the insertion of a ``tuning parameter'' $q$ in the CC approach, which 
we discuss in detail in section~\ref{sec:CC}.
We see that, upon suitable
tuning of this parameter, the ordinary differential
equations (ODEs), which emulate at a low-dimensional level the $\phi^6$ model, produce results that are in close qualitative (and even semi-quantitative) agreement  with the full numerical simulation of the governing partial differential equation (PDE).
{ It should be explicitly noted, however, that while the parameter
  $\epsilon$ of the potential will be an intrinsic parameter of the $\phi^6$
  PDE model
  considered here, $q$ is a phenomenological parameter inserted at the
  level of the CC ODE effective description of the system. Hence, $q$ is 
  a different type of parameter, arising during the coarse-graining of the
  original PDE into a set of ODEs.}
Finally, in section~\ref{sec:conc}, we summarize our results and
present some directions for future research. For completeness, we also include a relevant Appendix, in which we provide the details of the derivation of the complete (``unreduced'') CC theory, from which we  have derived the ``reduced'' CC description of our chosen $\phi^6$ model.

\section{Model Setup}\label{sec:model}

\begin{figure}[tbp]
\begin{center}
\includegraphics[width=0.6\textwidth]{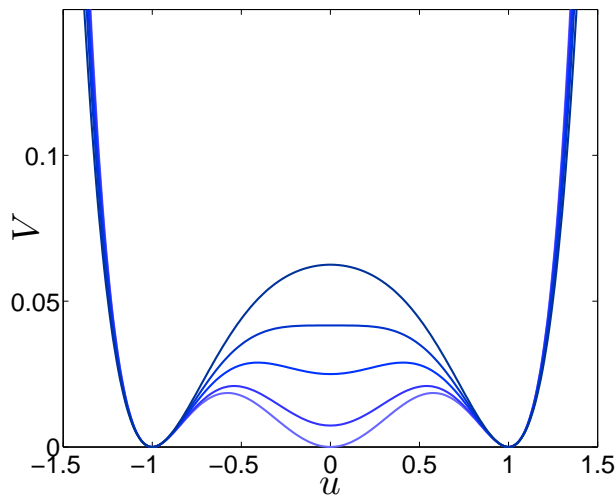}
\end{center}
\caption[]{The parametric $\phi^6$ potential given in Eq.~(\ref{eqn2}) as $\epsilon$ is varied from $0.01$ to $0.25$ to $0.5$ to $1/\sqrt{2}$ and to $1$ (lighter color curves to darker color curves, respectively), showing the transition from a triple well to a double well. The local minimum at $u=0$ disappears at $\epsilon=1/\sqrt{2}$.}
\label{fig:potential}
\end{figure}

\begin{figure}[tbp]
\begin{center}
\includegraphics[width=0.6\textwidth]{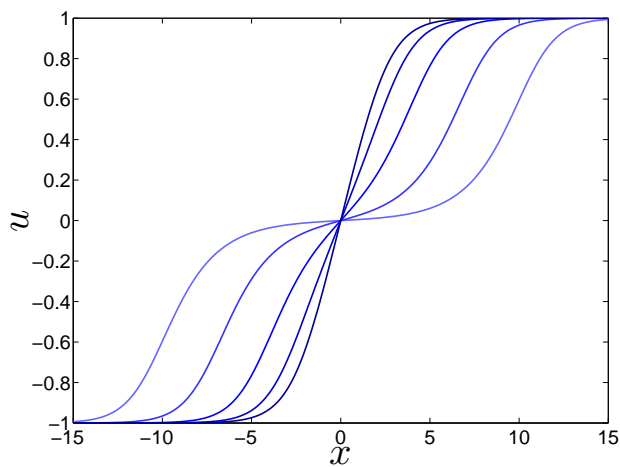}
\end{center}
\caption[]{The stationary solution, Eq.~(\ref{eqn3}), of our chosen $\phi^6$ model, as $\epsilon$ is varied from $0.01$ to $0.05$ to $0.2$ to $0.5$ and to $1$ (lighter color curves to darker color curves, respectively). Clearly, this solution is of the form of two kinks bound together.}
\label{fig:kinks}
\end{figure}

As discussed above, 
motivated by the work of Christ and Lee~\cite{lee1}, we explore a Klein--Gordon
field theory of the $\phi^6$ variety (setting $\phi=u$) in the form of the PDE:
\begin{equation}
u_{tt}=u_{xx} - V'(u),
\label{eqn1}
\end{equation}
where the potential is given by
\begin{equation}
V(u)=\frac{1}{8 (1+\epsilon^2)} (u^2 + \epsilon^2) (1- u^2)^2,
\label{eqn2}
\end{equation}
and it is depicted for several values of $\epsilon$ in Fig.~\ref{fig:potential}.
Notice that this is not precisely the potential proposed in~\cite{lee1},
however for simplicity we have set the parameter $\mu$ (controlling the
amplitude therein) and the parameter $g$ (controlling the value of
the uniform steady state) both to unity, for simplicity. The key
parameter remaining in the model is the parameter $\epsilon$, which is connected to the curvature of the potential and is
critical to understanding our spectral and dynamical observations below.
The stationary solution of the model in Eq.~\eqref{eqn1} is of the form~\cite{lee1}:
\begin{equation}
u=u_0(x)=\frac{{\rm sinh}\left(\frac{x}{2}\right)}{\sqrt{1+\epsilon^{-2}
+ \sinh^2\left(\frac{x}{2}\right)}}.
\label{eqn3}
\end{equation}
The remarkable feature of this solution, and the reason why it was
chosen as a candidate for a simplified one-dimensional ``bag'' model of quarks within hadrons, is
that it does {\it not} exist in the form of an isolated kink.
Instead, it takes the form of two kinks ``glued'' to each other in the
form of a bound state, as shown in Fig.~\ref{fig:kinks} for various choices of $\epsilon$. This is indeed reminiscent, at least
conceptually, of the setting of quarks whose total number is
conserved, but which possess infinite energy in isolation, while
certain multi-quark configurations thereof (in the form of bound
states) exist and possess finite energy. Topologically similar ``glued'' kinks were also discussed in 
\cite{sanati} in the context of a $\phi^6$ model relevant to first-order phase transitions in  materials science.

To make the connection with some of the recent works such as those
of~\cite{shnir1,danial}, let us touch upon the limit of $\epsilon \rightarrow 0^+$.
Interestingly, in this limit, our solution given in Eq.~(\ref{eqn3}) 
degenerates to a vanishing uniform steady state, while the solution 
explored in the work of~\cite{shnir1} is $u \propto \sqrt{(1+\tanh x)/2}$, 
connecting the fixed points (``vacuum'' states) $u=0$ and $u=1$. Furthermore, contrary to what
is the case herein, in the $\epsilon\to0$ limit there is no internal mode in the
linearization spectrum of the steady kink solution. In fact, the latter feature is the important
distinguishing trait that the authors of Ref.~\cite{shnir1} attribute to their work. In other words, while such
internal modes are absent in the kink linearization spectrum of the kink studied in \cite{shnir1},
the collisions between a kink and an antikink {\it still} feature
the resonance windows known, e.g., in the case of the standard
$\phi^4$ model~\cite{campbell}, wherein it was argued that the coupling to the 
internal mode of the $\phi^4$ kink leads to the resonance windows.

\begin{figure}[tbp]
\begin{center}
\includegraphics[width=0.6\textwidth]{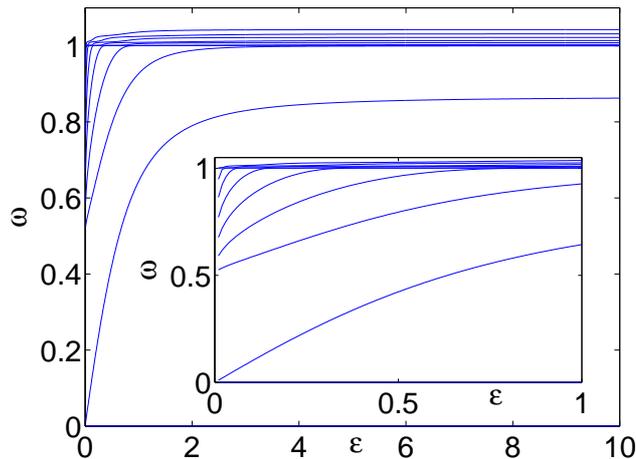}
\end{center}
\caption[]{The lowest (positive) eigenfrequencies of
the linearization problem given in Eq.~(\ref{eqn4}). The phonon band
(i.e., the continuous spectrum) starts at $\omega=1$, hence eigenfrequencies
below that critical point are bifurcating internal modes. One can
clearly distinguish (also through the inset) the progressive
increase in the number of internal modes as $\epsilon\to0^+$.}
\label{fig_ic1}
\end{figure}

On the contrary, a remarkable feature of the model discussed
herein, i.e., Eq.~(\ref{eqn1}), is that as the
curvature-controlling parameter $\epsilon$ is decreased, the number
of internal modes in the model continues to increase. In particular, 
an approximate argument in \cite{lee1} provides support for
the number of internal modes being of $\mathrm{O}(-\epsilon^{-1}\log\epsilon)$
as $\epsilon\to0^{+}$.  In the present work, we will
 systematically analyze the linearization 
spectrum of Eq.~(\ref{eqn1}) around the bound-state kink solution
given in Eq.~(\ref{eqn3}).  Specifically, we 
perturb the bound-state solution $u_0(x)$ given in Eq.~(\ref{eqn3}) as follows: 
\begin{equation}
u(x,t) = u_0(x) + \delta e^{i \omega t} \chi(x),
\end{equation} 
then we substitute this ansatz back into Eq.~(\ref{eqn1}) and linearize the problem to $\mathrm{O}(\delta)$. 
Next, we solve the resulting linear problem, namely,
\begin{equation}
-\omega^2 \chi= \chi'' -V''(u_0) \chi,
\label{eqn4}
\end{equation}
for the eigenfrequency-eigenvector pair $\big(\omega,\chi(x)\big)$.
In particular, in Fig.~\ref{fig_ic1}, the smallest
eigenvalues (eigenfrenquencies) of the linearization problem are shown.
In our current setting of Eq.~(\ref{eqn4}) with the potential
given in Eq.~(\ref{eqn2}), the phonon band (i.e., the continuous
spectrum of the problem) extends over the interval
$\omega\in(-\infty,-1) \cup (+1,+\infty)$. Hence, eigenfrequencies
below $\omega=1$ in the positive semi-axis of Fig.~\ref{fig_ic1}
correspond to internal modes associated with the kink.
Of course, in addition to these internal modes,
  there exists a mode at $\omega=0$, reflecting the translational
  invariance of the model. The remaining frequencies $\omega$ belonging to the point-spectrum of Eq.~(\ref{eqn4}) depend on $\epsilon$.

One natural limit worth considering is that of $\epsilon \rightarrow
\infty$. In that case, we revert to the standard, well-known 
$\phi^4$ model situation, which is known to have only one internal
mode at $\omega=\sqrt{3}/2$ (in the present units). This is clearly
captured accurately in the present numerical computations.
We now proceed to explain the picture ``shooting down'' from 
that special limit. In so doing, we identify a second 
internal mode. The latter seems to be asymptotically bifurcating
(as $\epsilon \rightarrow \infty$) from the band edge
(rather than stemming from a clear-cut, finite value of $\epsilon$).
However, additional modes progressively emerge in the point spectrum.
These bifurcations can be quantified as occurring at 
$\epsilon=0.856$, $\epsilon=0.356$, $\epsilon=0.15$, $\epsilon=0.063$
and $\epsilon=0.0286$ for the five subsequent ones, giving
rise to 7 discernible modes down to the case of $\epsilon=0.01$ 
explored here. Furthermore, it should be noted that the dynamics
of the first internal mode (the one that exists even as $\epsilon 
\rightarrow \infty$) appear to be in agreement with
the expectation of \cite[Fig.~2]{lee1} (in that work,
the additional modes were not systematically probed).
An additional observation worth making
  here concerns the limit of $\epsilon \rightarrow 0^+$.
  In this limit, as can already be discerned in Fig.~\ref{fig:kinks},
  the two kinks, constituting the kink of interest herein,
  ``part'' from each other, creating two separate
  kinks one connecting $-1$ to $0$ and one connecting
  $0$ to $+1$. These isolated (in the limit) kinks are
  ``individually'' translationally invariant. This
  rationalizes the approach of a second frequency (pair)
  to $\omega=0$ in this limit, reflecting the individual
  translational invariance of these two kinks.

These findings about the internal mode structure and its
variation with $\epsilon$ lead to some key questions.
In particular, it is relevant
to identify settings where there is predominantly a single
internal mode and compare/contrast them to ones where there are
multiple internal modes. For instance, taking $\epsilon=10$,
we have a picture bearing essentially a single internal mode.
Then, the collisions between a kink and an antikink should
bear the characteristics of two-bounce windows, three-bounce windows
and four-bounce windows similar to the intervals obtained in, e.g.,
\cite[Tables I-III]{anninos}. More modern ways of visualizing
the relevant data can be found in, e.g., \cite[Fig.~1]{shnir1}.
On the other hand, if we lower the value of $\epsilon$, say to 
$\epsilon=1$, where visibly two internal modes are present, we would expect kink collision phenomenology to be significantly modified. Subsequent
examinations of the cases of, say, $\epsilon=0.5$ and
then $\epsilon=0.25$ would enable the probing of cases
with three and four internal modes. Comparing/contrasting the
dynamical outcomes and collisional features of these different
cases promises to provide a systematic way to understand 
the role of internal modes in the kink dynamics. It is, thus, this task
that we embark upon next.

\section{Numerical Results}\label{sec:NR}

In this section we present numerical simulations of the collisions of the kink-antikink pair of Eq.~(\ref{eqn1}) for values of $\epsilon \in (0,20]$. {The kink solution is given by $u_0(x)$ in Eq.~(\ref{eqn3}), while the antikink is given by $u_0(-x)=-u_0(x)$ for the present case. The initial field configuration is taken to be of the form of a kink ($K$) plus an antikink ($\bar{K}$). To consider the collision of moving kinks and antikinks, the stationary solution given in Eq.~(\ref{eqn3}) is Lorentz boosted, so that
\begin{equation*}
u_{K}(x,t)= u_0\big(\gamma (x+x_0-vt)\big), \qquad u_{\bar{K}}(x,t)=-u_0\big(\gamma (x-x_0+vt)\big),
\end{equation*}
where $\gamma:=1/\sqrt{1-v^2}$, $x_0$ is where the kink is initially centered such that $2x_0$ is the initial separation between the kink and antikink, and $v$ is the velocity of the moving kink. (The velocity of the moving antikink is $-v$.) Then, the initial conditions for our simulations are given by
\begin{subequations}\begin{align}
u(x,0)  &=u_{K}(x,0) + u_{\bar{K}}(x,0)-1, \\
u_t(x,0)&=(u_{K})_{t}(x,0) + (u_{\bar{K}})_t(x,0),
\end{align}\label{KKbar-IC}\end{subequations}
as illustrated in Fig.~\ref{fig:kink_antikink}. 

\begin{figure}[tbp]
\begin{center}
\includegraphics[width=0.6\textwidth]{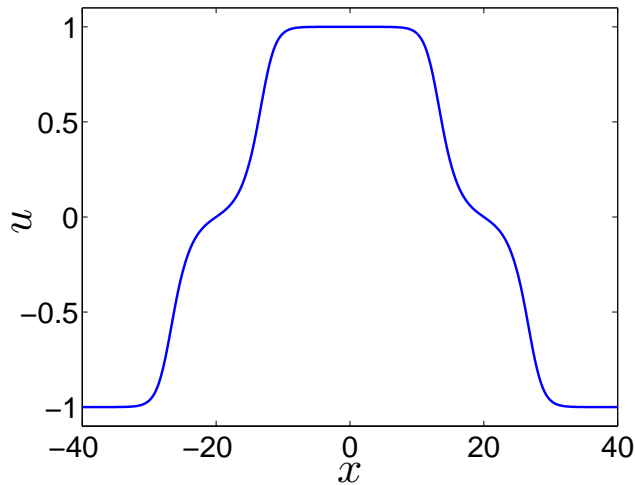}
\end{center}
\caption[]{An example of (stationary) kink-antikink configuration with $\epsilon=0.05$ and $x_0=20$.}
\label{fig:kink_antikink}
\end{figure}

To similate the collisions of kinks and antikinks, we discretized the continuous PDE (\ref{eqn1}) using central finite differences in space, which yields the following semi-discrete set of equations:
\begin{equation}
\ddot{u}_{m}=\frac{1}{(\Delta x)^2}(u_{m+1}-2u_m+u_{m-1})  - V'(u_m),  
\end{equation}
where over-dots denote time derivatives. For a sufficiently small $\Delta x$, typically $\Delta x=0.1$,  the continuous solution is well approximated by $u_m(t) \approx u(x_m,t)$, where $m=0, 1, 2, \hdots, 2N$ and $x_m=(m-N) \Delta x$. The spatial domain is taken to be the interval $[-N\Delta x, N \Delta x]$, where $N=699$ is used in our simulations. Then, we employ a fourth-order Runge--Kutta method for the evolution in time with a fixed time-step of $\Delta t = 0.001$.

As is well known, the governing PDE~(\ref{eqn1}) has Hamiltonian structure, therefore it conserves an energy (Hamiltonian) functional given by
\begin{align}
 H = \mathcal{T}(u;t) + \mathcal{V}(u;t) = \int_{-\infty}^{+\infty} \left(\frac{1}{2}u_t^2 + \frac{1}{2}u_x^2 + V(u)\right) \,\mathrm{d}x ,
\label{eq:H_par}
\end{align}
where the kinetic $\mathcal{T}$ and potential $\mathcal{V}$ energy contributions of the field, respectively, are
\begin{align*}
\mathcal{T}(u;t) &= \frac{1}{2} \int_{-\infty}^{+\infty} u_t^2 \,\mathrm{d}x,\\
\mathcal{V}(u;t) &= \int_{-\infty}^{+\infty} \left(\frac{1}{2} u_x^2 + V(u)\right) \mathrm{d}x.
\end{align*}
Since $\mathrm{d}{H}/\mathrm{d}t=0$, ${H}$ is a given constant for a chosen initial field configuration. In our simulations, the average value of ${H}$ is of $\mathrm{O}(1)$, while the deviations from the mean are (for the numerous examples we considered) no more than $10^{-13}$. Thus,  we have used energy conservation as a check on the validity of our numerical results.

We take the kink and antikink in the initial configuration (\ref{KKbar-IC}) to have equal and opposite initial velocities, $\pm v_\mathrm{in}$ (as can always be arranged in this translationally invariant field theory in the center-of-mass frame by a boost transformation of the kink and antikink). The possible outcomes of these collisions are either the so-called multi-bounce scenarios (including as a special case the single bounce one, where the kink and antikink separate immediately after a single collision) or a capture and formation of the so-called bion state~\cite{campbell,Campbell1}. In the former case, an $n$-bounce represents a scenario for which upon $n$ successive bounces ($n=1$, $2$, $3$, etc.) between the kink and antikink, they collect sufficient kinetic energy to escape each other's attraction and finally separate from each other asymptotically. Naturally, for a sufficiently large velocity $v_\mathrm{in}>v_\mathrm{c}$, for some critical velocity $v_\mathrm{c}$, the canonical scenario is that of $n=1$.  When these waves eventually separate, they escape with equal and opposite velocities $\pm v_\mathrm{out}$. On the flip side, when $v_\mathrm{in}$ tends to zero, it is natural to expect that the kinks do not possess enough kinetic energy to escape each other's attraction. In that case, they will form a very long-lived bound state termed a bion. The multi-bounce windows are interspersed between these two well defined limits (of large and small initial velocity $v_\mathrm{in}$). 

\begin{figure}[tbp]
\begin{center}
\subfigure[]{{\includegraphics[width=0.49\textwidth]{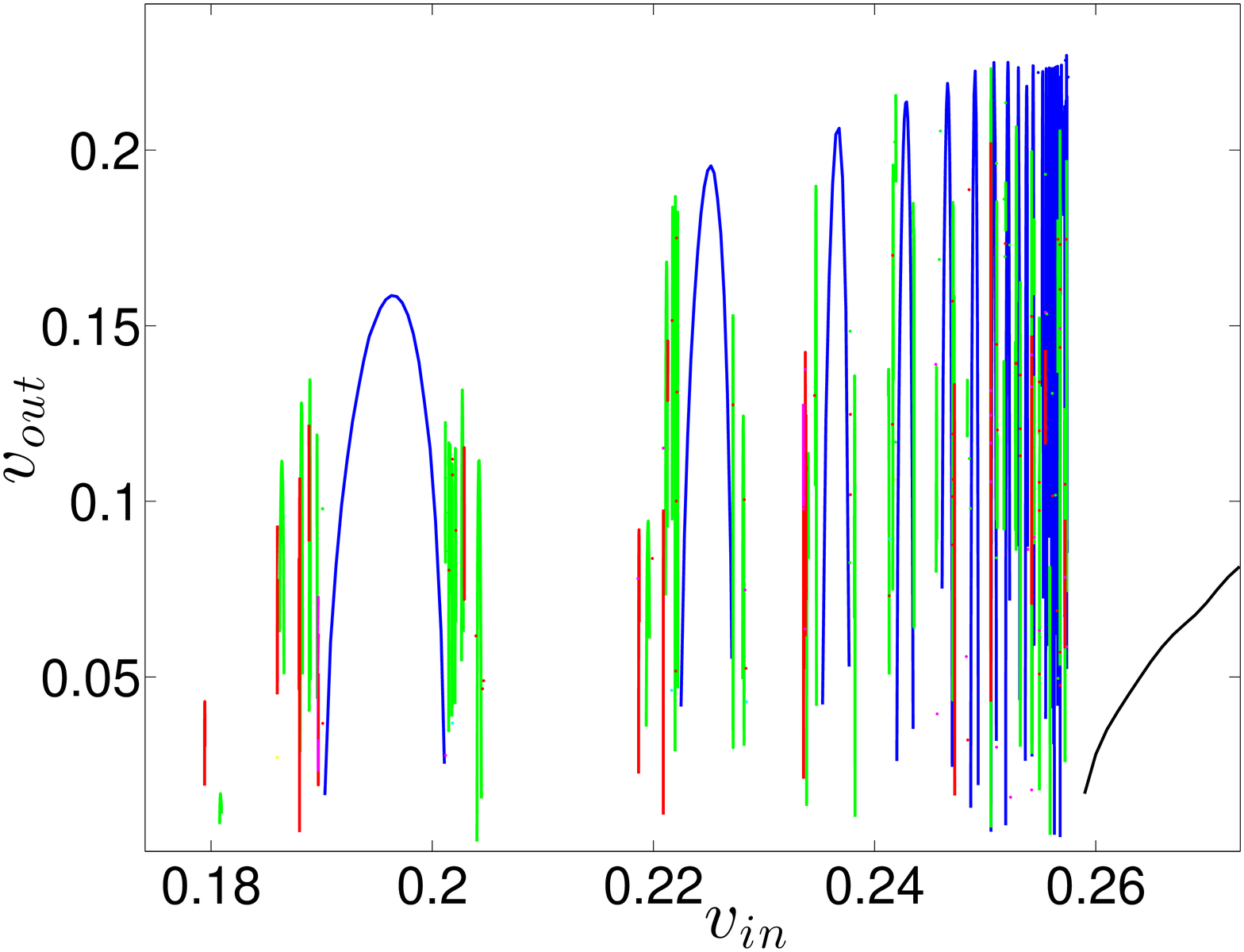}}}
\subfigure[]{{\includegraphics[width=0.49\textwidth]{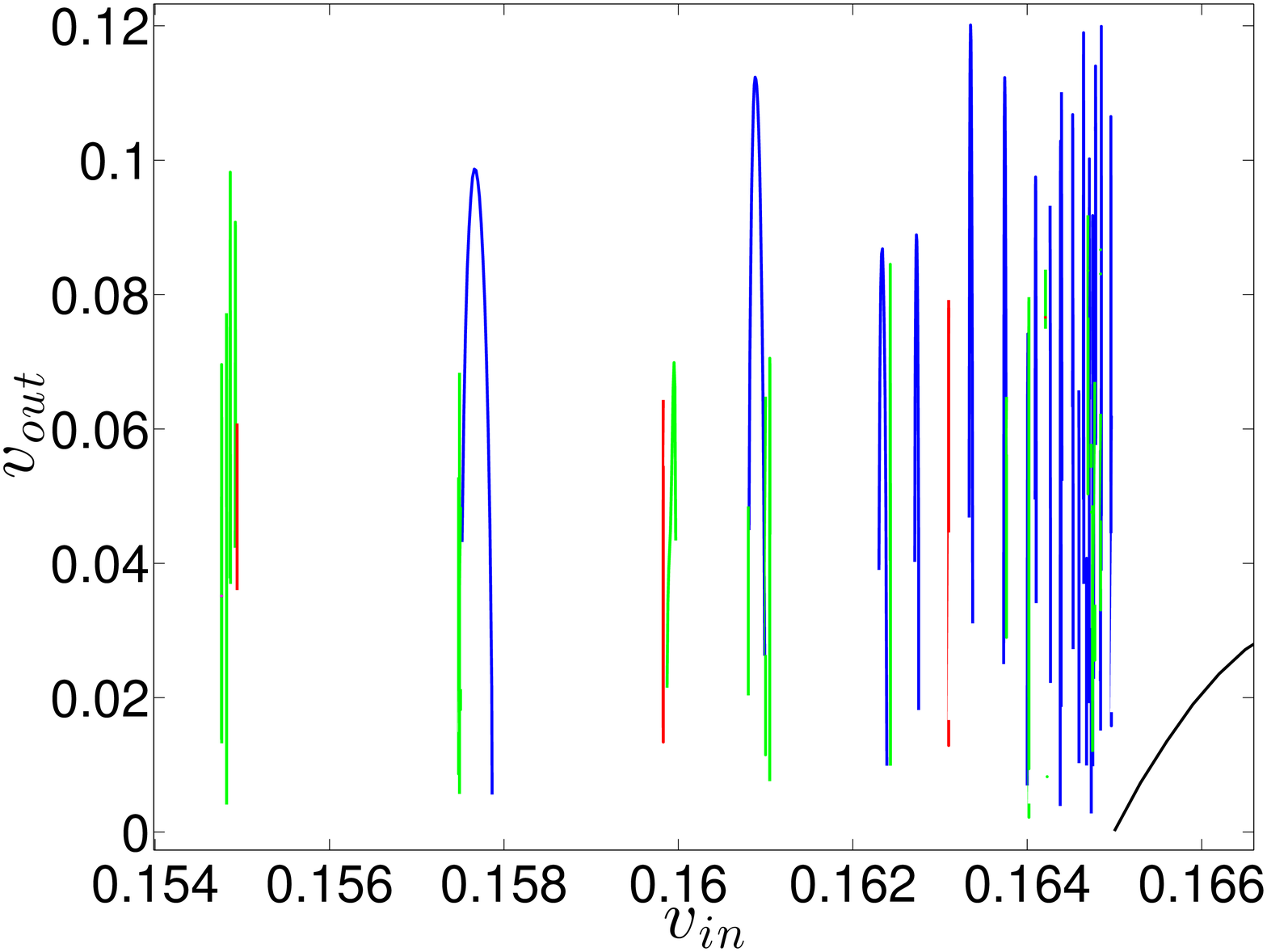}}}
\subfigure[]{{\includegraphics[width=0.49\textwidth]{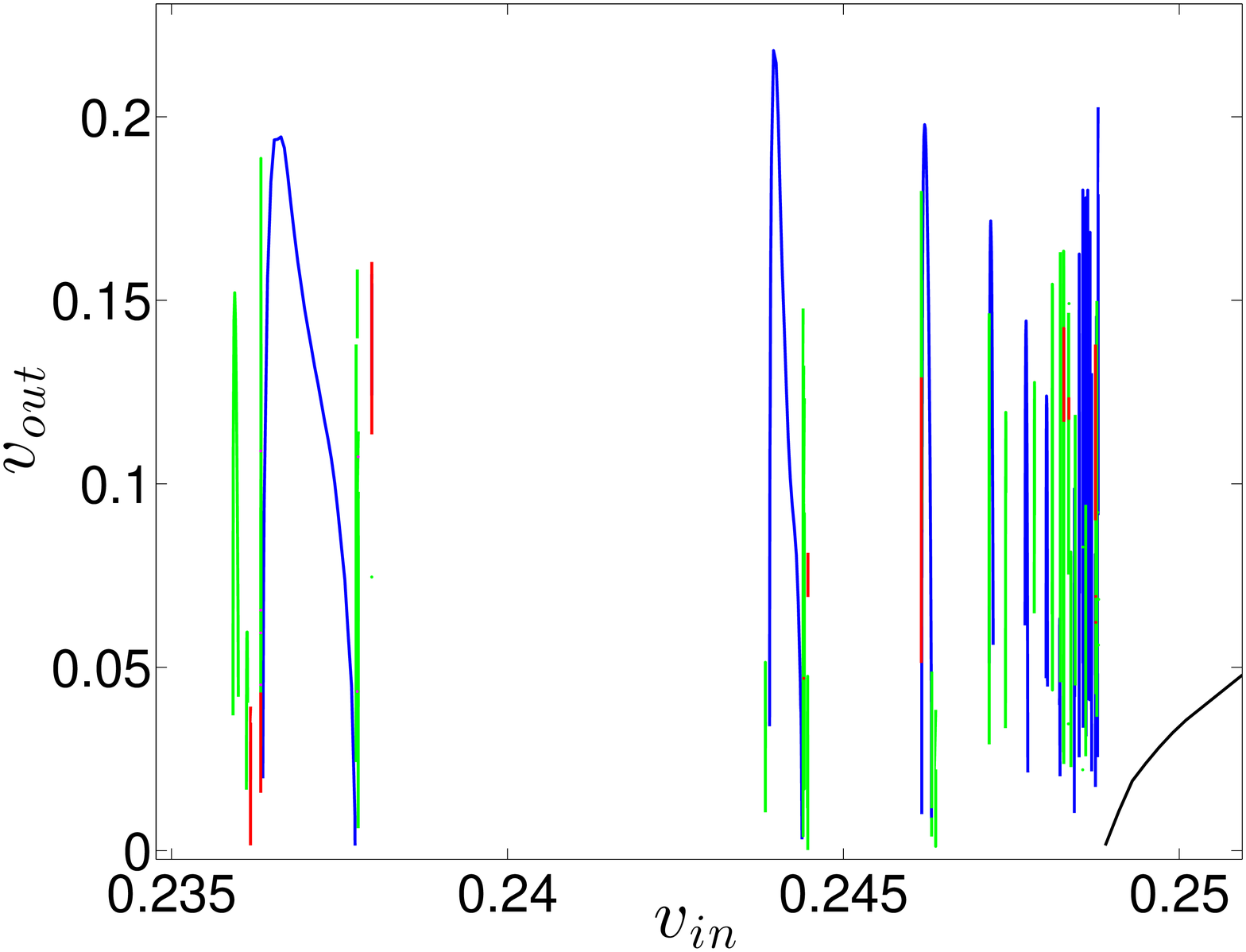}}}
\caption{The escape velocity $v_\mathrm{out}$ as a function of the initial velocity $v_\mathrm{in}$ when (a) $\epsilon=20$ (b) $\epsilon=1$  (c) $\epsilon=0.5$. Black represents a 1-bounce solution, blue represents a 2-bounce solution, green represents a 3-bounce solution, and red represents a 4-bounce solution.\label{vin_vout}}
\end{center}
\end{figure}

 Following the type of representation given in~\cite{goodman2},
  Fig.~\ref{vin_vout} shows how the escape velocity $v_\mathrm{out}$ depends on the initial velocity $v_\mathrm{in}$ when $\epsilon=20$, $\epsilon=1$, $\epsilon=0.5$
  (i.e., going from one internal mode to two and then to three). { Note that in order to calculate $v_\mathrm{out}$, we perfomed a linear interpolation of two points (position vs time) and calculated its slope. We made sure that the $t$ values were sufficiently large so that  $v_\mathrm{out}$ remained constant in time.}
  For bigger values of $\epsilon$ in Fig.~\ref{vin_vout}(a), as expected, we obtain results similar to \cite{goodman2,goodman}, since our $\phi^6$ model behaves like the classical $\phi^4$ model.  As $\epsilon$ becomes smaller, it gets harder to detect the intervals of $n$-bounce solutions, see Fig.~\ref{vin_vout}(b,c).  It can be
  observed that as $\epsilon$ becomes smaller, i.e., as a larger
  number of internal modes become available, remarkably the dynamics
  appears to become ``less complex''. In other words, there are fewer multi-bounce
  windows, and they extend over a narrower band of initial velocities
  $v_\mathrm{in}$. This transition is clear when going from panel
  (a) to panel (b) in Fig.~\ref{vin_vout}, and perhaps even more dramatically
  when going from panel (b) to panel (c). This feature
  appears to be consistent with the corresponding conclusions
  regarding a parametric double-well model in Ref.~\cite{simas}.
  The presence of multiple internal modes decreases the frequencies of the internal vibrations (with $\epsilon$).  Apparently, in addition, the presence of multiple internal modes pushes the multi-bounce windows closer to the critical
  velocity (cf.\ the analytical predictions for the windows, e.g., in~\cite{campbell,simas}), which also significantly disturbs the delicate resonance structure
  involved in multi-bounces. The combination of these effects appears
  to be responsible for the reduction in collisional complexity.

In Fig.~\ref{n-bounce}, we present $n$-bounce examples in the context of Eq.~(\ref{eqn1}) for $n=1, 2, 3, 4,5$ and $\epsilon=1$. {In the plots, $x_t$ is the approximate center of the kink (top) and antikink (bottom) as defined by their intersection with the $x$-axis, and $t$ represents  time in the PDE model. } Figure~\ref{n-bounce}(c) shows that for an initial velocity $v_\mathrm{in}=0.155$, the kink and antikink form a bound pair.  This is the bion state discussed above. The other limiting-case scenario is encapsulated in Fig.~\ref{n-bounce}(f), which shows a one-bounce scenario. This occurs when the initial velocity $v_\mathrm{in}$ is picked greater than the critical velocity $v_\mathrm{c}$, which is the velocity beyond which the kink and antikink with equal and opposite initial velocities collide once, then separate asymptotically. For initial velocities below this critical value, the kink and antikink are either captured or separate eventually after several bounces [see panels (e), (b), (d) and (a) of Fig.~\ref{n-bounce} for $n=2$, $3$, $4$ and $5$, respectively].
  
\begin{figure}[tbp]
\begin{center}
\subfigure[]{{\includegraphics[width=0.325\textwidth]{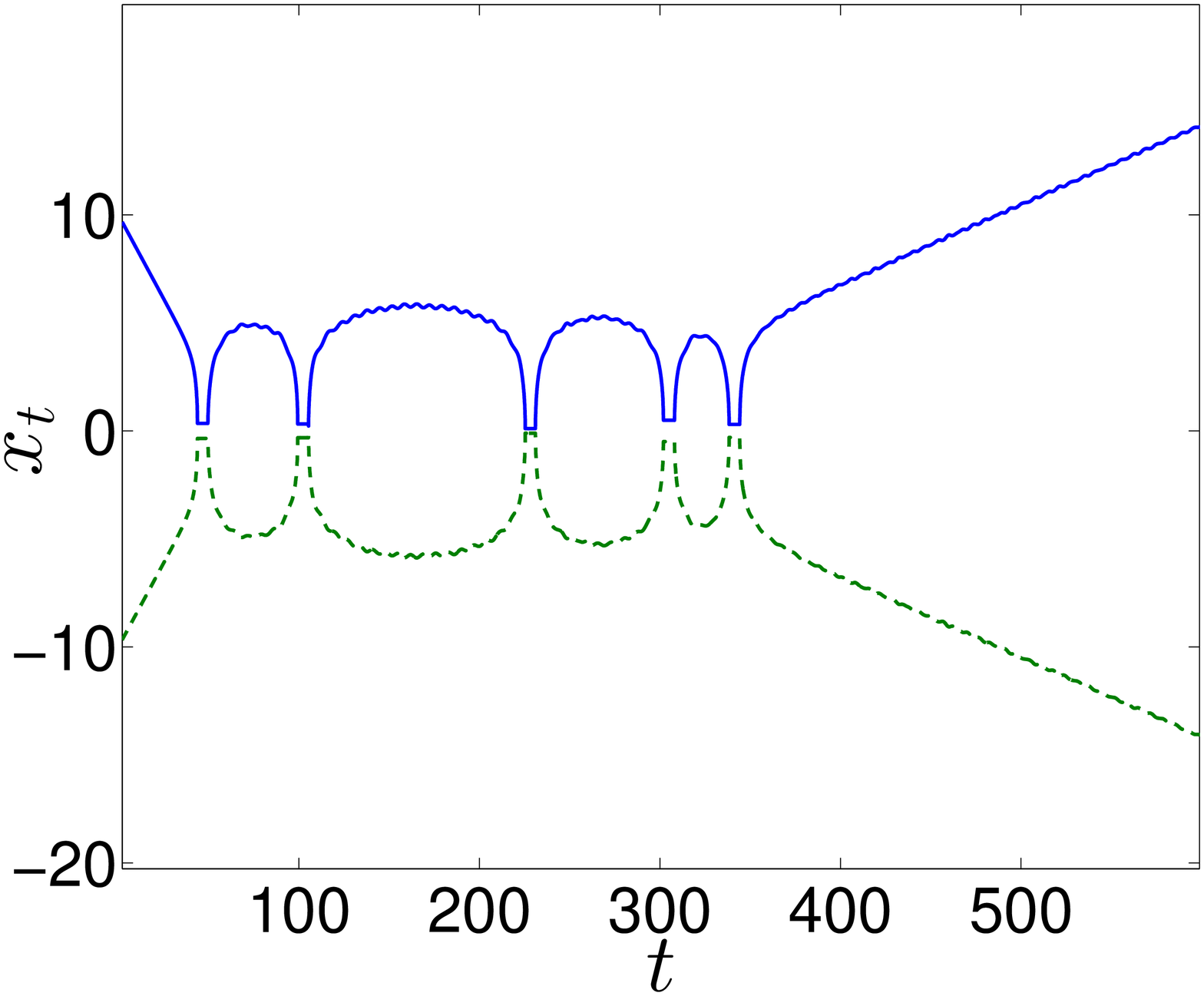}}}
\subfigure[]{{\includegraphics[width=0.325\textwidth]{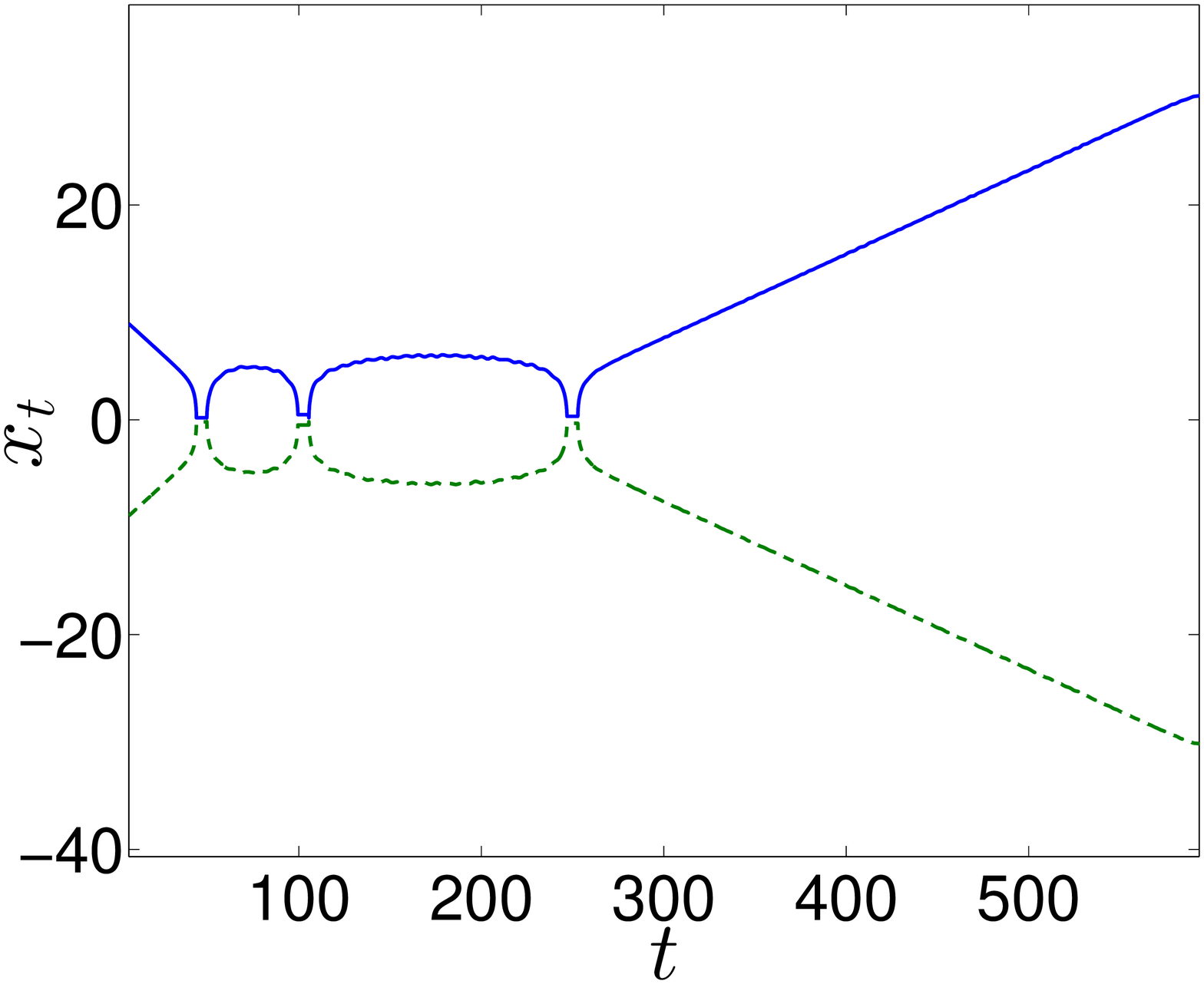}}}
\subfigure[]{{\includegraphics[width=0.325\textwidth]{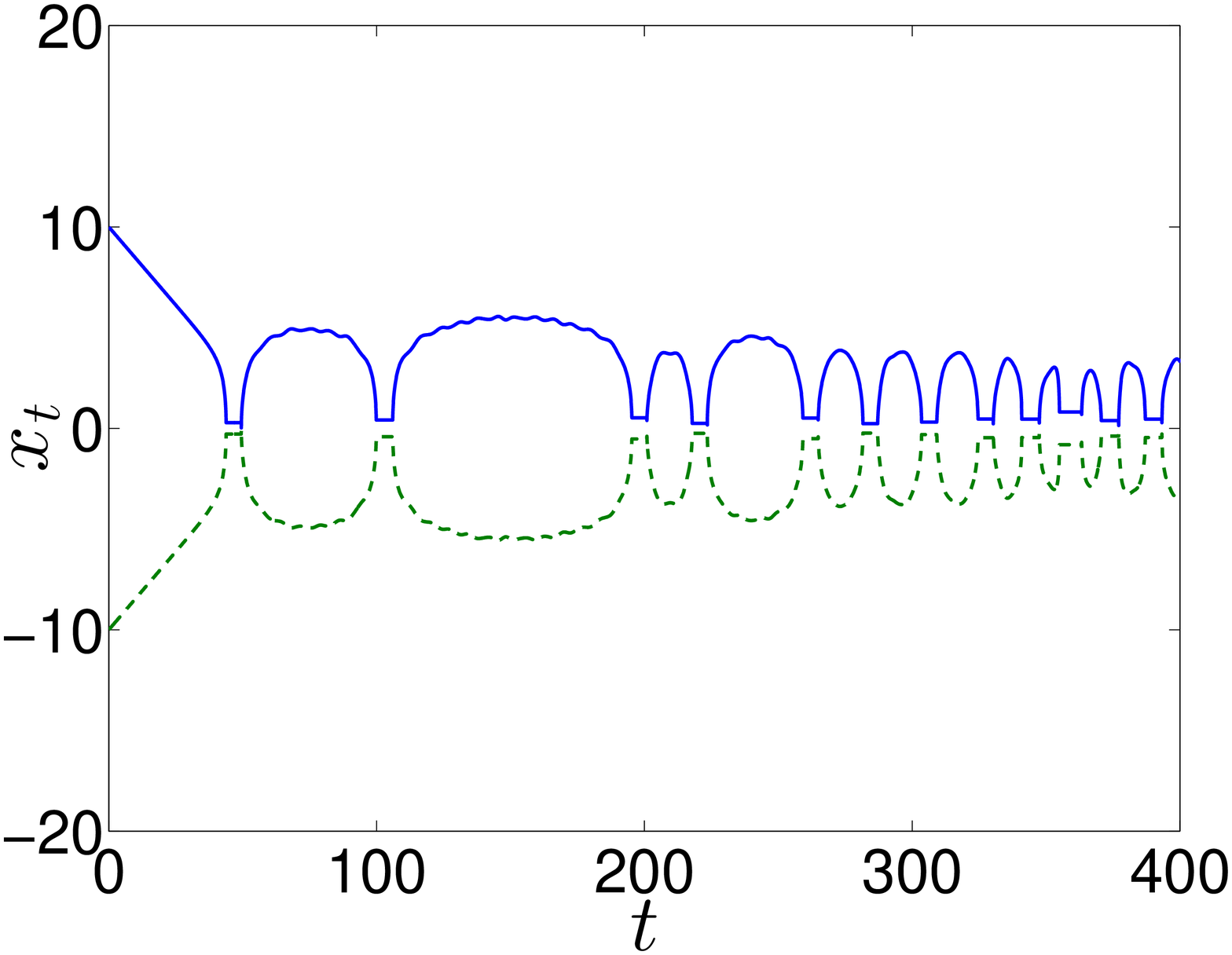}}}
\subfigure[]{{\includegraphics[width=0.325\textwidth]{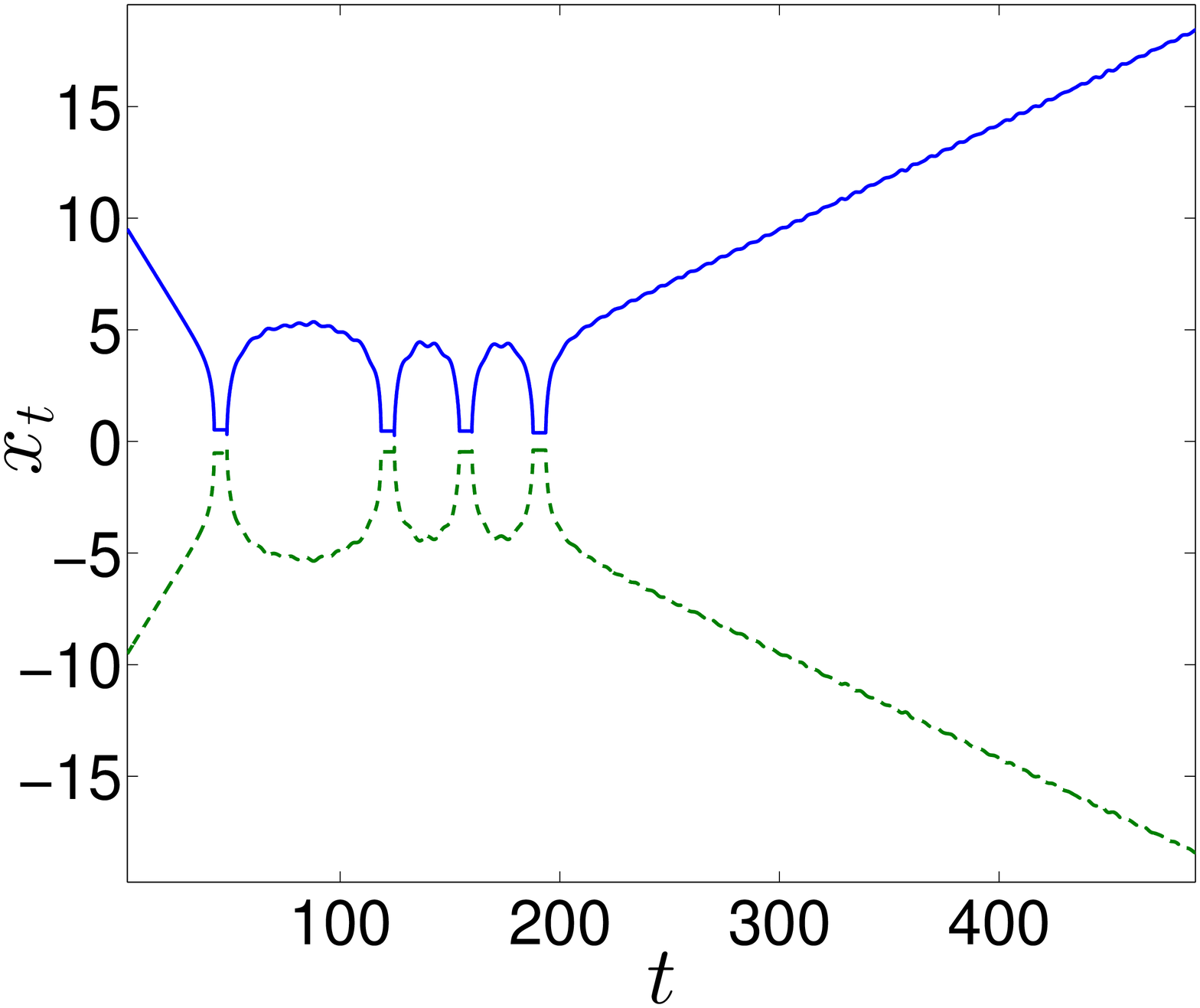}}}
\subfigure[]{{\includegraphics[width=0.325\textwidth]{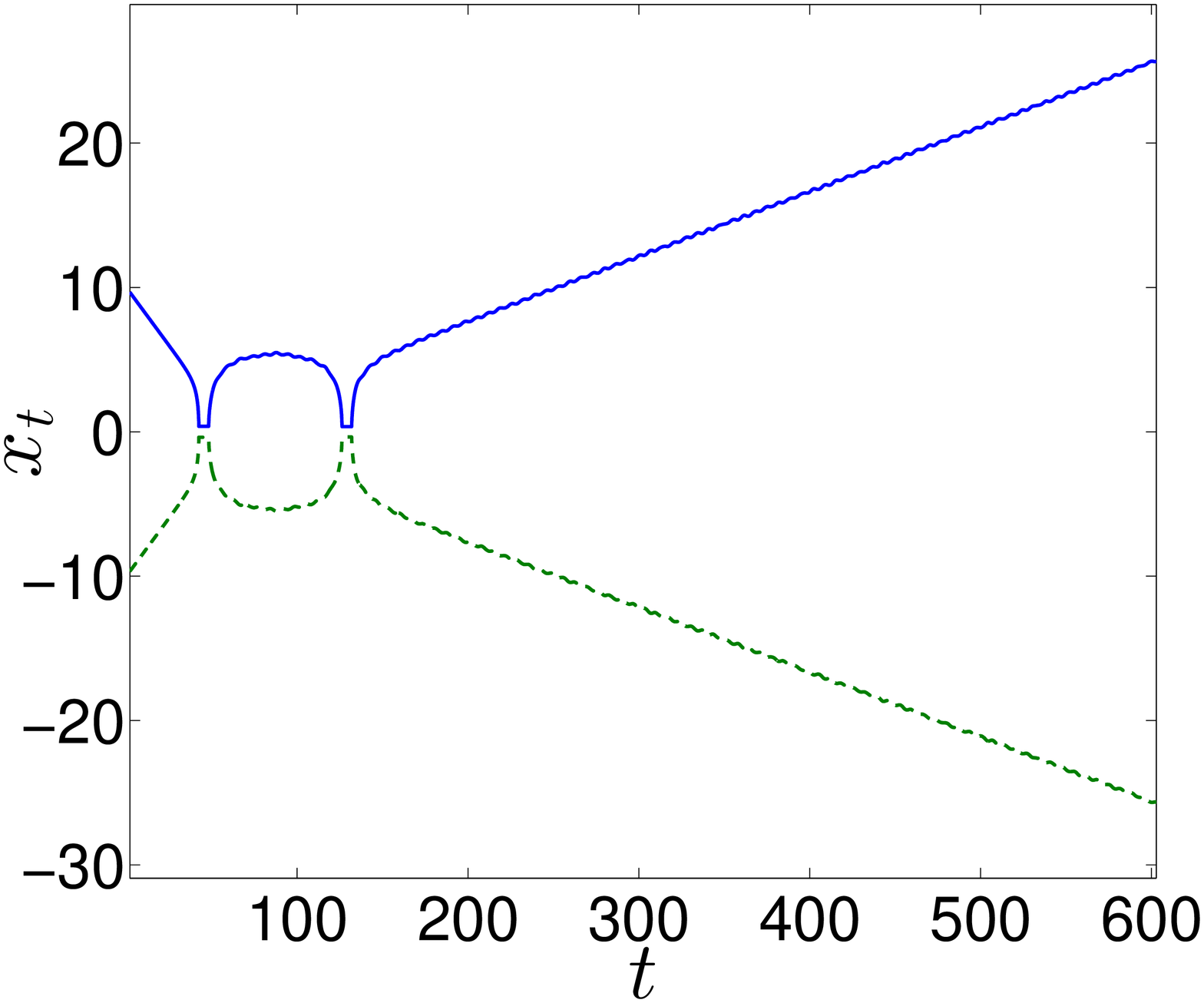}}}
\subfigure[]{{\includegraphics[width=0.325\textwidth]{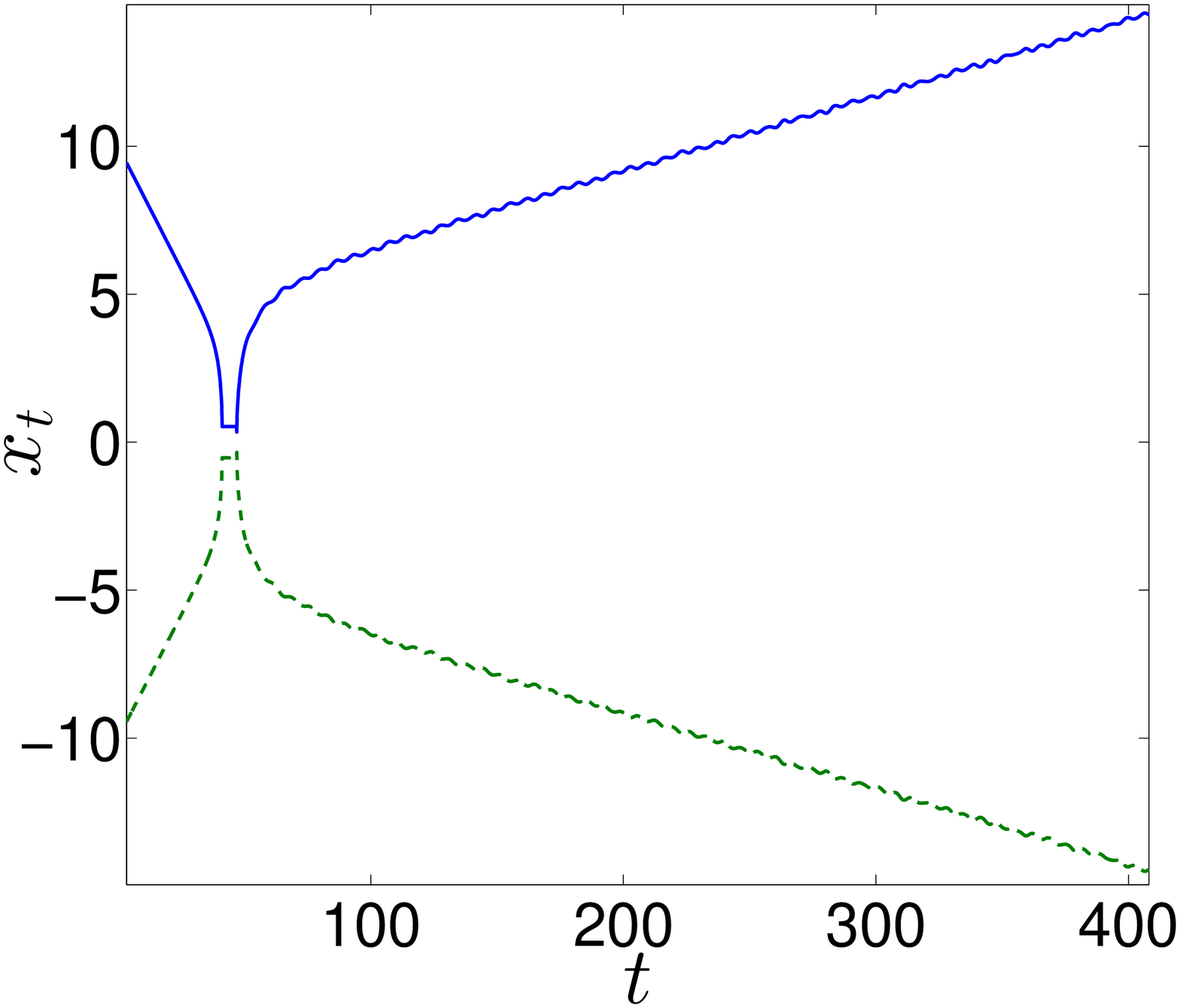}}}
\end{center}
\caption{Kink-antikink collisions when $\epsilon=1$. (a) A 5-bounce solution at $v_\mathrm{in}=0.154761850$. (b) A 3-bounce solution at $v_\mathrm{in}=0.154818$. (c) Capture at $v_\mathrm{in}=0.155$. (d) A 4-bounce solution at $v_\mathrm{in}=0.1598265$. (e) A 2-bounce solution at $v_\mathrm{in}=0.16081$. (f) A 1-bounce solution at $v_\mathrm{in}=0.1665$.}
\label{n-bounce}
\end{figure}

\begin{figure}[tbp]
\begin{center}
\includegraphics[width=0.55\textwidth]{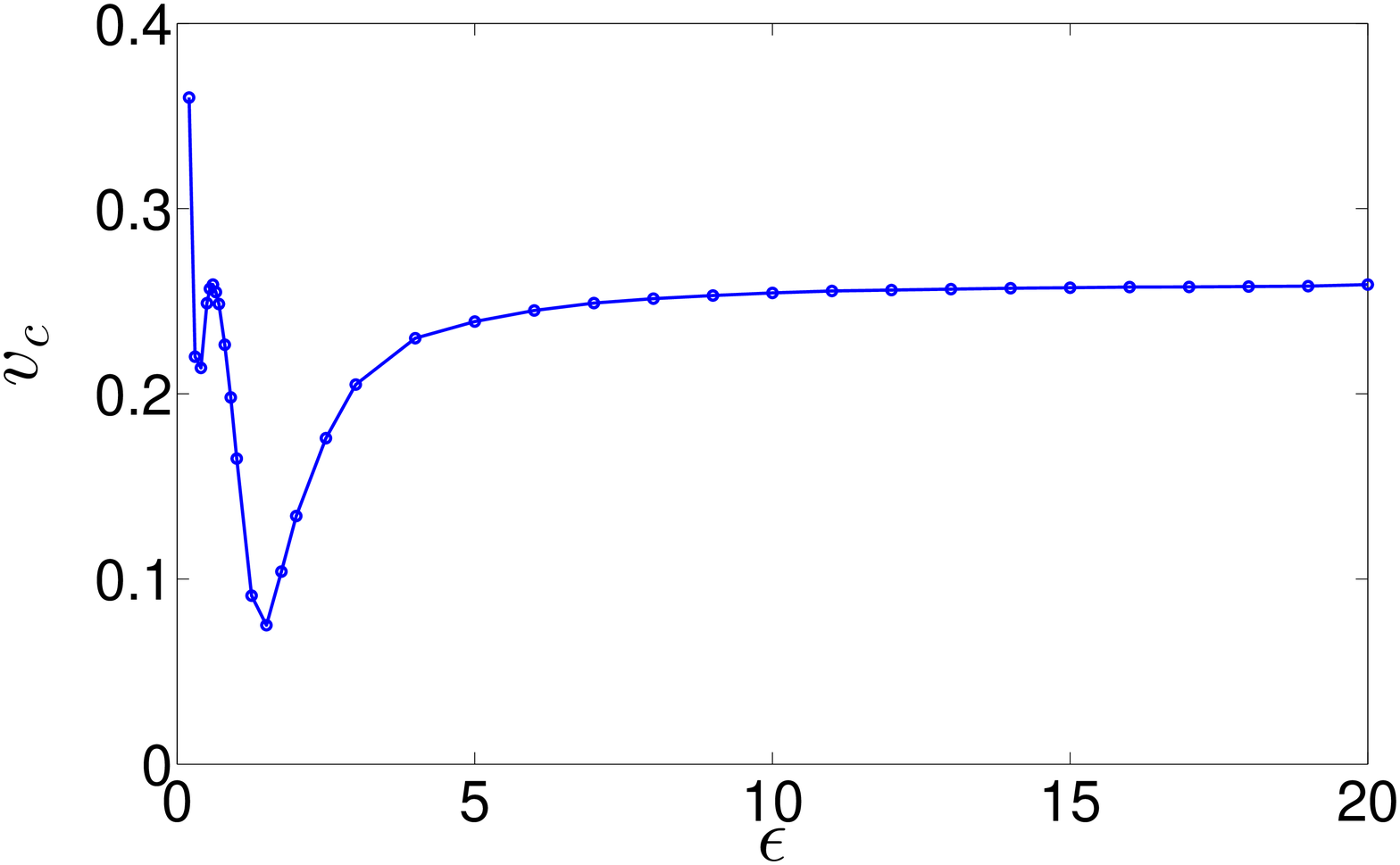}
\caption[]{The critical velocity $v_\mathrm{c}$ as a function of $\epsilon$. The numerical simulations show that $\epsilon \approx 1.5$ is a cutoff for the monotonic behavior of this function. For any $\epsilon$ greater than this cutoff value of $1.5$, we see that $v_\mathrm{c}$ increases monotonically with $\epsilon$. However, this is not the case for $\epsilon$ less than the cutoff value.}
\label{eps_vs_vc}
\end{center}
\end{figure}
  
Importantly, we also show the critical velocity $v_\mathrm{c}$ varies with $\epsilon$. Figure~\ref{eps_vs_vc} shows that for larger values of $\epsilon$, $v_\mathrm{c}$ increases monotonically with $\epsilon$. However, the situation changes for small values of $\epsilon$, and a non-monotonic trend involving oscillations appears to arise as $\epsilon\to0^+$. While this may be somewhat surprising in its own right, on the one hand a somewhat similar non-monotonic dependence has been identified as a function of the free parameter in the potential in the work of~\cite{simas} for a generalized double-well potential. On the other hand, we will see that the CC approach utilized below will be able to capture this phenomenology. It should also be noted that the sextic character  of the potential clearly becomes evident for $\epsilon < 1/\sqrt{2}$, when the third well emerges. For $\epsilon > 1/\sqrt{2}$, the potential features a local maximum at the origin (i.e., for $u=0$), appearing as a double rather than as a triple well potential as illustrated in Fig.~\ref{fig:potential}.

\section{Collective Coordinate Approach and Connection to the Numerical Results}\label{sec:CC}

In this section, we use the method of collective coordinates (CCs) to understand the kink-antikink interactions studied numerically in the previous section. Following the time-honored tradition of a wide variety of early works, we reduce the PDE (\ref{eqn1}) to a Hamiltonian dynamical system with two degrees of freedom: the kink position $X(t)$ and the magnitude of its internal mode $A(t)$.
However, we will follow the prescription of~\cite{weigel2}, given the problems described therein with regard to utilizing the ``standard'' CC methodology, which we have encountered as well. In particular, numerical integration of the full ODE system arising in the CC approach breaks down as $X(t)\to 0$. In order to resolve such problems, as proposed in \cite{weigel2}, we assume a colliding kink-antikink system with the following field configuration 
\begin{multline}
 u (x,t)=u_{0}\big(x+X(t)\big)-u_{0}\big(x-X(t)\big)-\frac{1}{2}\big[1+\tanh\big(qX(t)\big)\big]\\
+A(t)\Big[\chi_{\epsilon}\big(x+X(t)\big)-\chi_{\epsilon}\big(x-X(t)\big)\Big], 
 \label{kink-antikink}
\end{multline}
where $X(t)$, the half the distance between the kink and antikink, and $A(t)$, the amplitude of the internal mode perturbation, are the two degrees of freedom of the CC description. Here, as before, $u_0(x) $ is the stationary kink solution given in Eq.~(\ref{eqn3}). In addition, it should be explicitly stated here that our ansatz (\ref{kink-antikink}) utilizes a collective coordinate based on a linearized (internal) mode of the kink $\chi_\epsilon$ and its associated lowest (positive) eigenfrequency $\omega_{\epsilon}$ of
the linearized problem (\ref{eqn4}). Note that the $\epsilon$ subscript on $\chi_{\epsilon}(x)$ and $\omega_{\epsilon}$ reminds the reader that they both depend on $\epsilon$ and (typically) must be calculated numerically by solving Eq.~(\ref{eqn4}).  The factor $q$ in Eq.~\eqref{kink-antikink} was introduced in~\cite{weigel2}
to avoid some of the pathologies of the standard CC reductions; see the relevant discussion therein. The numerical value of $q$ chosen depends on $\epsilon$, as will be explained below. Finally, we have neglected shape changes of the kinks in the CC ansatz Eq.~\eqref{kink-antikink} under the assumption that their velocities remain small for most of the duration of their interaction.

Next, we recall that the Lagrangian density for Eq.~(\ref{eqn1}) is 
\begin{equation}
\mathcal{L}=\frac{1}{2}u_{t}^{2}-\frac{1}{2}u_{x}^{2}-V(u),
\label{lag_density}
\end{equation}
where $V(u)$ is given in Eq.~(\ref{eqn2}). Then, the Lagrangian is  
\begin{equation}
L= \int_{-\infty}^{+\infty} \mathcal{L}\,\mathrm{d}x = \int \left[\frac{1}{2}u_{t}^{2}-\frac{1}{2}u_{x}^{2}-V(u) \right] \mathrm{d}x,
\label{Lagrangian}
\end{equation}
{where henceforth all integrals are understood to be over $x\in(-\infty,+\infty)$.}
Substituting Eq.~(\ref{kink-antikink}) into Eq.~(\ref{Lagrangian}) yields a lengthy expression. This ``unreduced'' Lagrangian is given in the Appendix as Eq.~(\ref{Lagrangian_expanded}) for completeness. { Following~\cite{goodman2}}, we work with a reduced effective Lagrangian that captures the fundamental features:
\begin{equation}\label{lag1}
L(X,\dot{X}, A, \dot{A}) = \big(M_0+I(X)\big)\dot{X}^2-U(X)+
\dot{A}^2-\omega_{\epsilon}^2A^2+2F(X)A,
\end{equation}
which follows from the general derivation given in the Appendix. Here, overdots denote time derivatives. As shown in the Appendix, $F$ in Eq.~(\ref{lag1}) is given by
\begin{multline}
 F(X)=\int \big[ V^{\prime }(u_{0}(x+X))-V^{\prime }(u_{0}(x-X))\big]\chi
_{\epsilon}(x+X)\,\mathrm{d}x \\
-\int V^{\prime }\left(u_{0}(x+X)-u
_{0}(x-X)-\frac{1}{2}\big[1+\tanh(qX)\big] \right)\chi
_{\epsilon}(x+X)\,\mathrm{d}x
\label{eq:F}
\end{multline}
and characterizes the interaction between the kink's translational and internal
modes. Meanwhile, 
\begin{equation}
 M_0=\int [u_{0}^{\prime }(x)]^{2}\,\mathrm{d}x
\label{eq:M0}
\end{equation}
is the rest mass of the kink, and
\begin{multline}
 I(X)= \int u_{0}^{\prime }(x+X)u
_{0}^{\prime}(x-X)\,\mathrm{d}x -\frac{q}{2}\int \big[ u_{0}^{\prime}(x+X)+u_{0}^{\prime}(x-X)\big] \sech^2(qX)\,\mathrm{d}x \\
+\frac{q^2}{8}\int \sech^4(qX)\,\mathrm{d}x
\label{eq:I}
\end{multline}
is an effective mass associated with the kink-antikink interaction and the interaction of both the kink and antikink with the field correction proposed by~\cite{weigel2}. Here, primes denote differentiation with respect to a function's argument. Finally,
\begin{multline}
 U(X)=\int \frac{1}{2}\big[ u_{0}^{\prime }(x+X)-u
_{0}^{\prime }(x-X) \big]^{2}\,\mathrm{d}x\\
+\int V\left(u_{0}(x+X)-u_{0}(x-X)-\frac{1}{2}\big[1+\tanh(qX)\big]\right)\mathrm{d}x  
\label{eq:U}
\end{multline}
represents an effective potential energy landscape due to the kink-antikink interaction.
\begin{figure}
\begin{center}
\includegraphics[width=\textwidth]{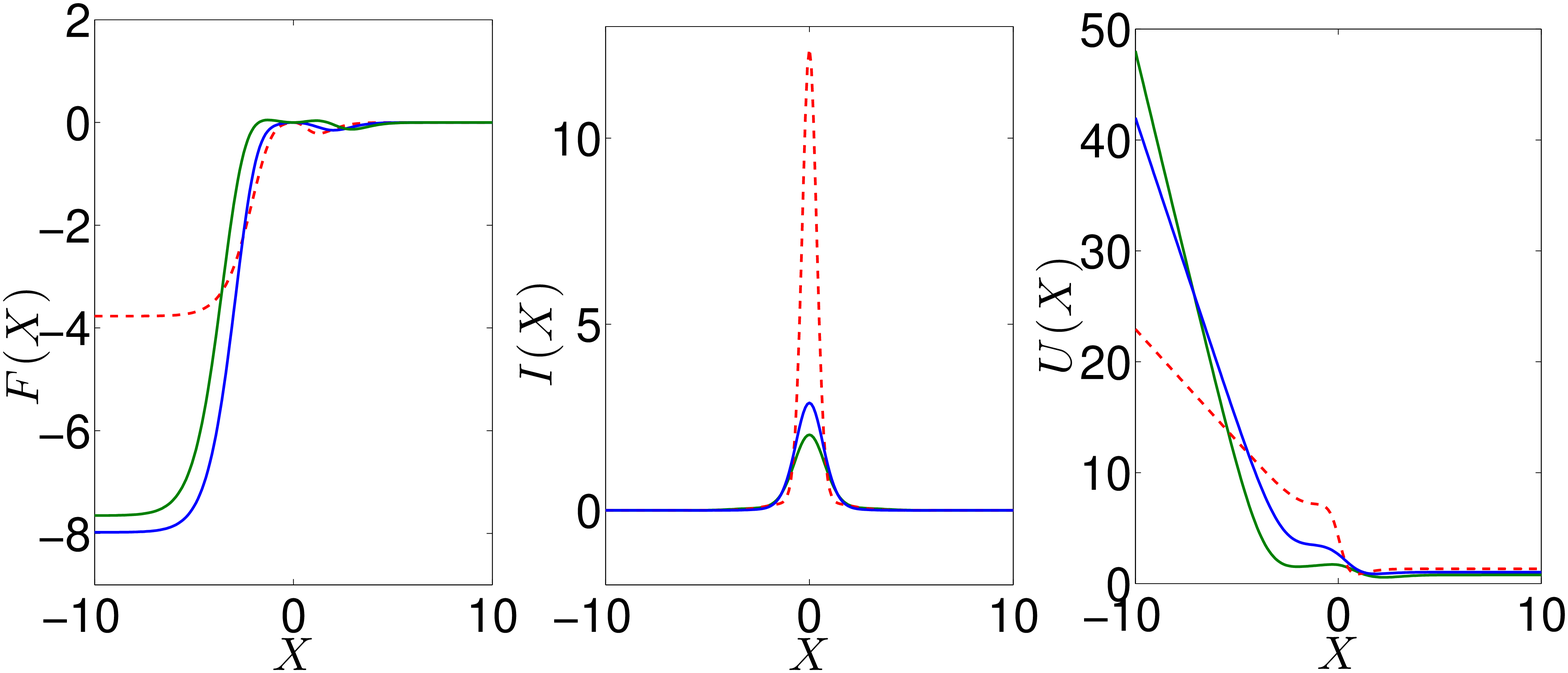}
\caption[]{Plots of the functions $F(X)$ (left), $I(X)$ (middle)
  and $U(X)$ (right) as a function of the position $X$ of the kink's
center, evaluated through the method of collective coordinates, as discussed in the text. Blue corresponds to $\epsilon=1$, $q=0.7115$. Green corresponds to $\epsilon=0.5$, $q=0.6196$. Dashed red corresponds to $\epsilon=20$, $q=1.3917$. }
\label{combined_coefficients}
\vskip-0.4cm
\end{center}
\end{figure}
Typically, $F(X)$, $U(X)$ and $I(X)$ must be evaluated by numerical quadratures.
In Fig.~\ref{combined_coefficients}, we show the functions $F(X)$, $U(X)$ and $I(X)$ for $\epsilon=0.5, 1 $ and $20$.
We can see that, while some of these functions roughly maintain the
same qualitative characteristics as the parameter $\epsilon$ is varied, their quantitative variations
are rather significant.
We have picked a specific $q$ value for each $\epsilon$. The process of choosing $q$ will be explained below.
Our goal is to investigate how the internal mode is excited (i.e., how $A$ evolves) in the kink-antikink collision process.

The Euler--Lagrange equations corresponding to Eq.~(\ref{Lagrangian}) are
\begin{subequations}\begin{align}
\frac{\mathrm{d}}{\mathrm{d}t}\left(\frac{\partial L}{\partial \dot{A}}\right) &=\frac{\partial L}{\partial A},\\ \frac{\mathrm{d}}{\mathrm{d}t}\left(\frac{\partial L}{\partial \dot{X}}\right) &= \frac{\partial L}{\partial X}.
\end{align}\label{EL-basic}\end{subequations}
Substituting Eq.~(\ref{lag1}) for $L$ in Eqs.~(\ref{EL-basic}) yields
\begin{subequations}\begin{align}
\ddot{A} &= -\omega_{\epsilon}^2A+F(X),\\
\big(2M_0+2I(X)\big)\ddot{X} & =-I'(X)\dot{X}^2-U'(X)+2F'(X)A.
\end{align}\label{red-ode}\end{subequations}
We solve this second-order system of ODEs, subject to the initial conditions $X(0)=x_0$ and $\dot{X}(0)=v_\mathrm{in}$, where $x_0$ is the initial half-distance between the kink and the antikink, and $v_\mathrm{in}$ is the initial velocity of the kink. We use MATLAB's built-in fourth-order Runge--Kutta variable-step size solver {\tt ode45} with built-in error control. We also derive the Euler--Lagrange equations corresponding to the ``unreduced'' Lagrangian (\ref{nonreduced}) and list all the formul\ae\ for the coefficients in the Appendix. The expressions involving integrals which are $X$-dependent in Eqs.~(\ref{red-ode}) are computed by numerical integration. 

As mentioned earlier, as $\epsilon \to \infty$, the
$\phi^6$ model considered herein converges to the $\phi^4$ model. Unfortunately, as discussed in \cite{weigel,weigel2}, the CC results for the $\phi^4$ model presented in \cite{sugiyama} contain some misprints. In addition, the reduced
system considered therein and in followup works neglects products of and higher-order terms in $A(t)$ and $X(t)$. When the CC formul\ae\ are augmented to include these terms, the results obtained by the  ODE system from the CC method do not agree as well with the results of solving the PDE (\ref{eqn1}) numerically. Specifically, a problem arises when $X$ is
very close to 0. In order to overcome these difficulties, in \cite{weigel2}, a modification of the field configuration that includes the terms involving the parameter $q$ was proposed:
\begin{multline}\label{q1}
 u(x,t)=u_{0}\big(x+X(t)\big)-u_{0}\big(x-X(t)\big)-\tanh(qX)\\
+A(t)\Big[\chi\big(x+X(t)\big)-\chi\big(x-X(t)\big)\Big].
\end{multline}
The latter is to be compared to the the ``standard'' field configuration introduced in \cite{sugiyama}, which would take the form
\begin{equation}\label{original}
 u(x,t)=u_{0}\big(x+X(t)\big)-u_{0}\big(x-X(t)\big)-1+A(t)\Big[\chi\big(x+X(t)\big)-\chi\big(x-X(t)\big)\Big],
\end{equation}
where $\chi(x)=\frac{\sqrt{3}}{2}\tanh(\frac{x}{2}) \sech(\frac{x}{2})$.

{In \cite{weigel2}, two models were studied, one is the $\phi^4$ model that is equivalent to our $\phi^6$ model in the limit as $\epsilon \rightarrow \infty$, and the second is the $\phi^6$ model corresponding to $\epsilon=0$ in our model. In \cite{weigel2} it was suggested that a different CC field ansatz be used for each model: (\ref{q1}) for the $\phi^4$ model,  and (\ref{kink-antikink}) for the $\phi^6$ model. Our numerical computations for the model considered herein suggest that for small values of $\epsilon$, a better { match} between the ODE and PDE results is obtained when ansatz (\ref{kink-antikink}) is used. For larger values of $\epsilon$, ansatz (\ref{q1}) gives a better match.} In \cite{weigel2}, the value of $q$ was chosen such that the escape velocity of the kink and antikink obtained by the CC approach matches with the escape velocity obtained from the numerical simulations of the PDE. However, then the number of collisions does not match in many cases. 

Our aim in the present work is to pick $q$ values such that the ODE results match the  PDE results to the fullest extent possible in all respects. In order to find the optimal $q$ for each $\epsilon$, we solve the second-order system of ODEs (\ref{red-ode}). The $q$
 value that makes the ODE and the PDE results match when $v_\mathrm{in}=v_\mathrm{c}$ becomes our optimal choice of $q$. Picking $q$ in this way allows us to better match the number of bounces predicted by the reduced-ODE model and the PDE. This ``optimal'' $q$ value is presented in Fig.~\ref{q_vs_eps} as a function of $\epsilon$, where we observe a mostly monotonic dependence of $q$ on $\epsilon$. {The fitted function (obtained from a standard numerical fitting routine for the range of $\epsilon$ shown in Fig.~\ref{q_vs_eps}) is given by
\begin{equation}\label{fitteddata}
 q(\epsilon)=\frac{1.356\epsilon^2-1.704\epsilon+3.179}{\epsilon^2-1.609\epsilon+4.742}.
\end{equation}
For instance, for $\epsilon=0.5$, the critical velocity is $v_\mathrm{c}=0.2489$ (recall Fig.~\ref{eps_vs_vc}). Solving the reduced-ODE system~(\ref{red-ode}) numerically, {using MATLAB's {\tt ode45} with the relative tolerance $10^{-5}$ and absolute tolerance $10^{-5}$}, we find that the best agreement between the CC approach and the PDE results occurs when $q=0.6196$.

\begin{figure}[tbp]
\begin{center}
\includegraphics[width=0.6\textwidth]{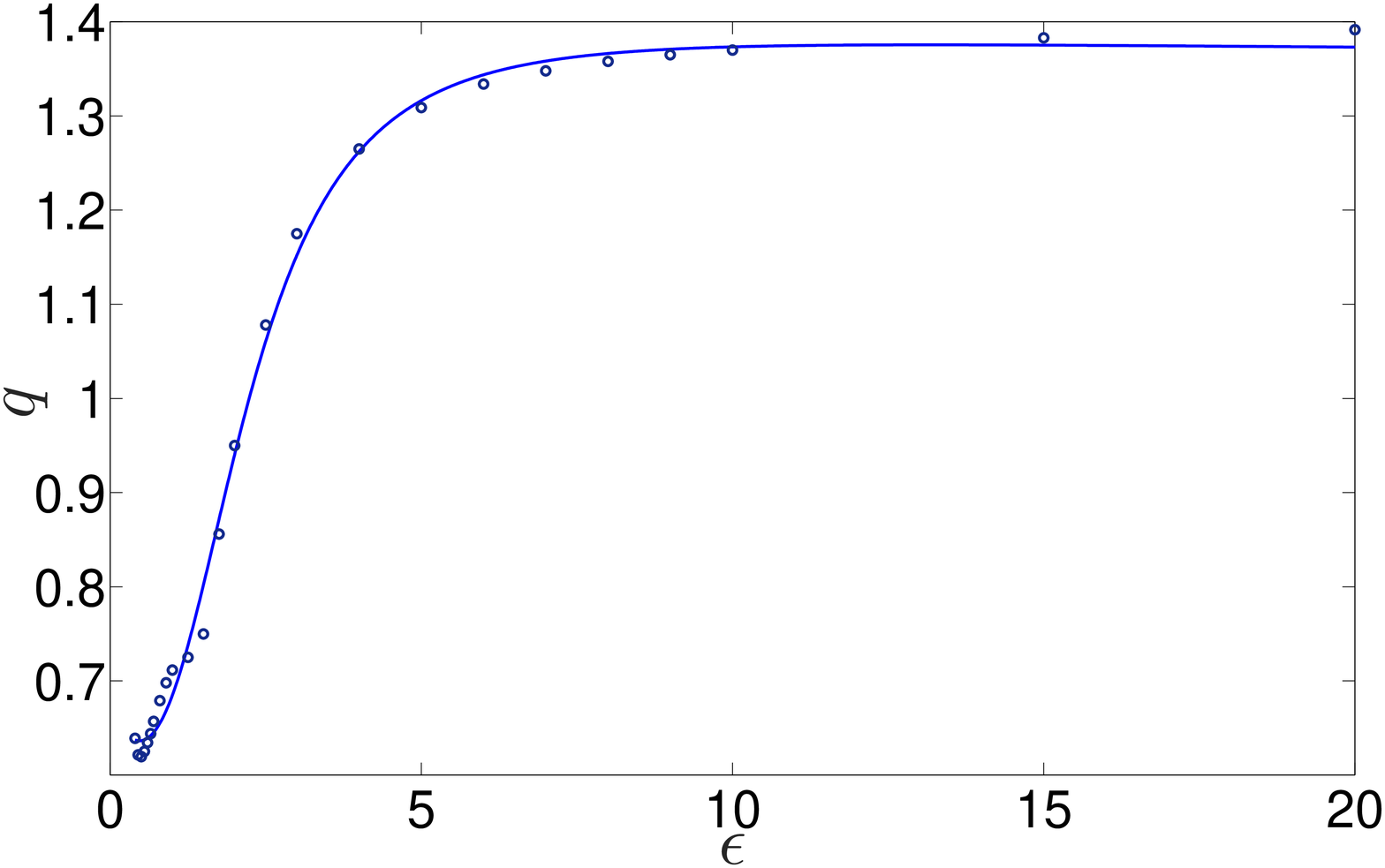}
\caption[]{Dependence of the CC ansatz tuning parameter $q$ on the curvature-controlling model parameter $\epsilon$. The data points and the corresponding fitting curve [given in Eq.~\eqref{fitteddata}] represent the relation between the ``optimal'' value of $q$ (under the notion of optimality defined in the text) and our $\phi^6$ model's free parameter $\epsilon$.}
\label{q_vs_eps}
\end{center}\end{figure}

\begin{figure}
\begin{center}
\includegraphics[width=\textwidth]{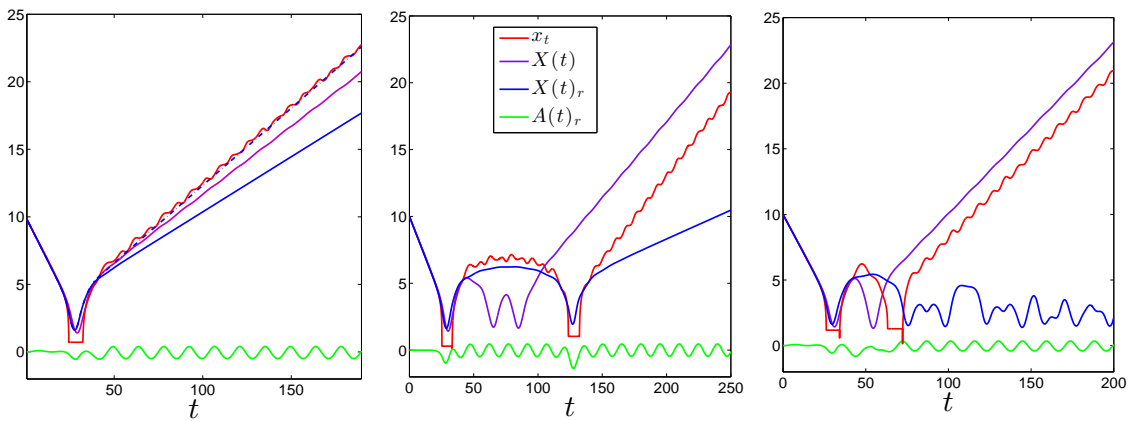}
\caption[]{Effect of shape mode(s) and (multi-)bounce windows for $\epsilon=0.5$, $q=0.6196$, $v_\mathrm{in}=0.26$ (left), $v_\mathrm{in}=0.24771$ (middle),  $v_\mathrm{in}=0.2372$ (right). In each panel, the curves represent the position of kink's center versus time obtained from: PDE (red), full-ODE (purple), reduced-ODE (blue), and the amplitude of the internal mode for the reduced-ODE (green). The blue dashed curve (left) corresponds to the reduced-ODE model with $q$ perturbing to be $0.6155$ in order to obtain a better match.}
\label{eps05_fig}
\vskip-0.4cm
\end{center}
\end{figure}

In Fig.~\ref{eps05_fig}, we show how the kink and antikink centers move in time, by using the CC-based reduced-ODE (\ref{red-ode}) (blue, solid), CC-base full-ODE  (\ref{ode-full}) (purple) and the PDE (\ref{eqn1}) (red). These results are based on the optimal value of $q$, which was computed as described above, thus differently from \cite{weigel2}. In the left panel, we additionally show the effects of slightly perturbing the value of $q$ so as to obtain a much better match between the reduced-ODE (blue, dashed) and the PDE (red). Each panel in Fig.~\ref{eps05_fig} shows that the results obtained by solving the full-ODE system (\ref{ode-full}) are in some ways better than the ones obtained by solving the reduced-ODE system (\ref{red-ode}). This is the case for most of our simulations. For instance, in the right panel of Fig.~\ref{eps05_fig}, we see that reduced-ODE and PDE results do not match at all. However, the number of bounces and escape velocities predicted by the full-ODE system do match the PDE. In the middle panel of Fig.~\ref{eps05_fig}, the reduced-ODE and the PDE results approximately match for a while but then the kink escape velocities disagree between the two approaches. On the other hand, the number of bounces of the full-ODE system and PDE-system do not match but their escape velocities do. Note that the full-ODE system is less sensitive to a change in $q$.

The results obtained by using the collective coordinates method show that for any $\epsilon\geq 0.3$, there exists a $q>0$ such that any $n$-bounce solution that solves the PDE (\ref{eqn1}) can be approximated. In order to obtain a better match, a very small perturbation of $q$ (within the range of $\pm 0.005$) is
sufficient; of course, on the flip side, these results suggest
the sensitivity of the comparison to the precise value of
$q$. However, the gross features of the collisions for any given
$\epsilon$ can be obtained by a particular value of $q$ based on
the monotonic correspondence given above in Eq.~(\ref{fitteddata}).
The numerical results obtained in Section \ref{sec:NR} show the relation between the critical velocity $v_\mathrm{c}$ and $\epsilon$ as in Fig.~\ref{eps_vs_vc}. The CC
method can capture this complex, non-monotonic
dependence as shown in Fig.~\ref{ode_verification}. However, when $0<\epsilon<0.3$, the CC results are not as accurate as the numerical solutions of the PDE. This feature, however,
can be rationalized on the basis of the large number of internal
modes that emerge as $\epsilon \rightarrow 0$, whose intricate effects
on the dynamics are not captured by the coarse-grained (yet already
rather complicated at the level of equations of motion) CC ansatz
described above. 
\begin{figure}[h!]\begin{center}
\includegraphics[width=0.8\textwidth]{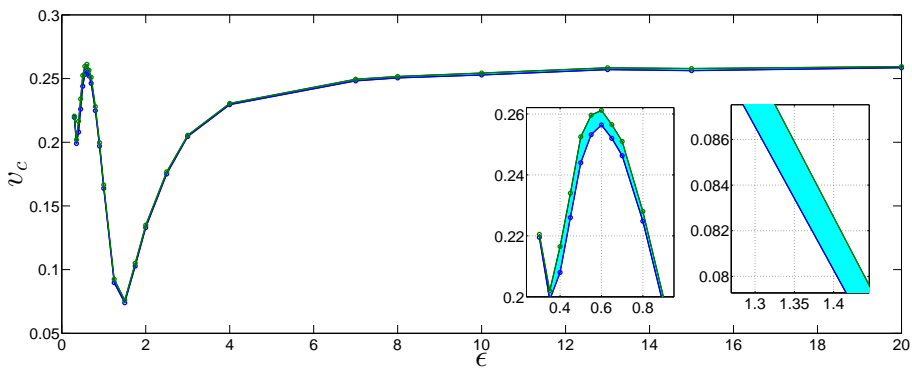}
\caption[]{The critical velocity $v_\mathrm{c}$ as a function of $\epsilon$ found via the method of collective coordinates. The plot shows a region between two curves. The inset plots are magnified views. If $q$ is the optimal value for a given $\epsilon$, then for values in the interval $q \pm 0.005$, $v_\mathrm{c}$ should lie in the shaded region. }
\label{ode_verification}
\end{center}\end{figure}

\section{Conclusions and Future Challenges}\label{sec:conc}

In the present work, we have revisited an intriguing mono-parametric
variant of the $\phi^6$ nonlinear field theory. This variant has numerous appealing features,
including the existence of an exact analytical solution, a
smooth variation of the potential from a triple well ($\phi^6$ type) to a double well ($\phi^4$ type),
the controllable emergence of progressively many internal modes
(as the parameter $\epsilon$ is reduced, recall Fig.~\ref{fig:potential}), among others. These features led
us to reconsider the collisions between a kink and an
antikink more generally. Such collisions were found to possess atypical features
such as: (a) progressively narrower and less complex (in their
structure) multi-bounce windows and (b) a non-monotonic dependence of
the critical velocity of the kink and antikink for a single bounce
event as a function of the potential parameter $\epsilon$; this led us to (c) consider
a modified form of the collective coordinates (CCs) approach,
based on~\cite{weigel2}, which enables us to capture qualitatively
and even semi-quantitatively the corresponding dynamics using a 
two-degrees-of-freedom formulation with a tunable parameter $q(\epsilon)$.

Our results are important for scalar field theory models \cite{makhankov} 
and more broadly for first-order phase transitions \cite{gufan,planes}. 
Naturally, these findings pave the way for a number of interesting
studies in the future.  Motivated by the findings of both the present work
 and that of~\cite{simas}, it is becoming especially relevant
to understand, from a qualitative perspective,
the modification of the multi-bounce windows
and the non-monotonic dependence of the critical velocity on
parameters smoothly deforming the model's potential. Hopefully, the collective
coordinates method used here will provide an avenue for further
analysis (e.g., in the spirit of~\cite{goodman}) in this direction.

Another question is whether models with more collective coordinates
are relevant, and whether it is possible to identify tangible
ways in which these additional CCs come into play, as the parameters
of the models are varied (and such modes bifurcate). While the methodology developed in~\cite{weigel2} appears to work
well overall for our problem (one could argue that it can be made/tweaked
to work even extremely well), it still perhaps lacks a solid
theoretical (conceptual) foundation, at least as far as the choice of
the tunable parameter $q$ is concerned.
These are important questions that we
believe will open new avenues in a problem that was, arguably, long thought to have been
definitively addressed. As such, we believe that these questions are certainly worthwhile
of further study and relevant results will be accordingly reported
in future publications.

\appendix
\section{Derivation of the Collective Coordinates Effective Lagrangians} 

We define $u_{\pm}:=\pm u_0(x\pm X(t))$ and $\chi_{\pm}:=\pm \chi_{\epsilon}(x\pm X(t))$, hence  $u'_{\pm}:=\pm u_0'(x\pm X(t))$ and $\chi'_{\pm}:=\pm \chi'_1(x\pm X(t))$. Then, Eq.~(\ref{kink-antikink}) becomes
\begin{equation} \label{kink-antikink_2}
u(x,t)=u_+ + u_{-} - \frac{1}{2}\big[1+\tanh(qX)\big] + A(\chi_{+}+\chi_{-}).
\end{equation}
Substituting Eq.~(\ref{kink-antikink_2}) into Eq.~(\ref{Lagrangian}) yields 
\begin{multline}
 L=\int  \left\{\frac{1}{2}\left[\left(u'_{+}-u'_{-}-\frac{q}{2} \sech^2(qX)\right)\dot{X}+\dot{A}(\chi_++\chi_-)+A\left(\chi'_+-\chi'_-\right)\dot{X}\right]^{2}\right\}\mathrm{d}x\\
-\int  \left\{\frac{1}{2}\left[ \left( u'_{+}+u'_{-}\right) +A\left(\chi'_++\chi'_-\right)\right]^{2} \right\}\mathrm{d}x-\int V(u)\,\mathrm{d}x.
\label{Lagrangian_expanded}
\end{multline}
For $V(u )$, we first write $u =u_{a}+u_{b}$ where $u
_{a}=u_++u_{-}-\frac{1}{2}[1+\tanh(qX)]$ and $u_{b}=A(\chi_{+}+\chi_{-}) $. Then, using a (finite) Taylor series, we get%
\begin{equation}
\begin{aligned}
 V(u ) &= V(u_{a}+u_{b})\\
 &= V(u_{a})+V^{\prime }(u_{a})u_{b}+\frac{V^{\prime \prime }(u_{a})}{2!}u_{b}^{2}+\frac{
V^{^{\prime \prime \prime }}(u_{a})}{3!}u_{b}^{3}+\frac{V^{(iv)}(u_{a})}{4!}u_{b}^{4} \\
 &\phantom{=} +\frac{V^{(v)}(u_{a})}{5!}u_{b}^{5} +\frac{V^{(vi)}(u_{a})}{6!}u_{b}^{6}.
\end{aligned}
\end{equation}
Since $V$ is a sixth degree polynomial, we have no higher terms in the Taylor series expansion.  Writing $V$ in this way allows us to keep track of the higher-order terms involving $A$ more easily.

Now, the CC Lagrangian (\ref{Lagrangian_expanded}) takes the ``unreduced'' form
\begin{equation}\label{nonreduced}
\begin{aligned}
 L(X,\dot{X}, A, \dot{A}) &= a_0(X)+a_1(X)A+a_2(X)A^2+a_3(X)\dot{X}^2 
 +a_4(X)\dot{A}^2+a_5(X)\dot{A}\dot{X}\\
 &\phantom{=}+a_6(X)A\dot{A}\dot{X}+a_7(X)A\dot{X}^2+a_8(X)A^2\dot{X}^2+a_9(X)A^3+a_{10}(X)A^4\\
 &\phantom{=}+a_{11}(X)A^5+a_{12}(X)A^6.
 \end{aligned}
\end{equation}
Writing the effective CC Lagrangian in this way highlights the order of each term though it is simply a formal manipulation to introduce the set of $X$-dependent coefficients $\{a_i\}_{i=1,2,\hdots,12}$. The formul\ae\ of the coefficients in Eq.~(\ref{nonreduced}) are given below. Note that they are all functions of $X(t)$. Since $\chi_{\epsilon}(x)$ is not known explicitly for every value of $\epsilon$, the coefficients are presented in integral form. These coefficients are calculated via numerical quadratures. Note that in the derivation of these coefficients, we repeatedly use the fact that a shifted function has the
same integral on an infinite interval as the unshifted one. For example, $
\int [u_{+}(x)]^{2}\,\mathrm{d}x=\int [u_{0}(x)]^{2}\,\mathrm{d}x=\int [u_{-}(x)]^{2}\,\mathrm{d}x$,
all of which are constants.

First, 
\begin{multline}
 a_0(X)=-\int \frac{1}{2}\big[ u_{0}^{\prime }(x+X)-u_{0}^{\prime }(x-X) \big]^{2}\,\mathrm{d}x\\
-\int V\left(u_{0}(x+X)-u_{0}(x-X)-\frac{1}{2}\big[1+\tanh(qX)\big]\right)\mathrm{d}x,
\label{eq:U}
\end{multline}
and $U(X)=-a_0(X)$ as given in Eq.~(\ref{eq:U}).

Next,
\begin{equation*}
 a_1(X)=\int -\left( u_{+}^{\prime }+u_{-}^{\prime }\right) \left(\chi
_{+}^{\prime }+\chi_{-}^{\prime }\right)-V^{\prime }\left(u_{+}+u
_{-}-\frac{1}{2}\big[1+\tanh(qX)\big]\right)(\chi
_{+}+\chi_{-})\, \mathrm{d}x.
\end{equation*}
Applying integration by parts on the first term of the last equation, we get
\begin{equation*}
 a_1(X)=\int \left( u_{+}^{\prime \prime }+u_{-}^{\prime \prime}\right) (\chi_{+}+\chi_{-})-V^{\prime }\left(u_{+}+u_{-}-\frac{1}{2}\big[1+\tanh(qX)\big]\right)(\chi_{+}+\chi_{-})\,\mathrm{d}x.
\end{equation*}
Using the fact that $u_{0}$ is the steady-state solution of Eq.~(\ref{eqn1}), we obtain
\begin{multline*}
a_1(X)=\int  \big[ V^{\prime }(u_{+})-V^{\prime }(u_{-})\big](\chi_{+}+\chi_{-})\\
-V^{\prime }\left(u_{+}+u_{-}-\frac{1}{2}\big[1+\tanh(qX)\big]\right)(\chi_{+}+\chi_{-})\, \mathrm{d}x.
\end{multline*}
Finally, by symmetry $X\rightarrow -X$, we get
\begin{align*}
 a_1(X) &=2\int \left\{ \big[ V^{\prime }(u_{+})-V^{\prime }(u_{-})\big] - V^{\prime }\left(u_{+}+u_{-}-\frac{1}{2}\big[1+\tanh(qX)\big]\right)\right\}\chi_{+}\, \mathrm{d}x.
\end{align*}
Note that $\displaystyle F(X)=a_1(X)/2$ as given in Eq.~(\ref{eq:F}).

Next,
\begin{multline}
a_2(X) = \int -\frac{1}{2}\left(\chi_{+}^{\prime }+\chi_{-}^{\prime
}\right)^{2} \,\mathrm{d}x \\ -\int \frac{1}{2}V^{\prime \prime }\left(u_{+}+u
_{-}-\frac{1}{2}\big[1+\tanh(qX)\big]\right)(\chi_{+}+\chi_{-})^{2}\,\mathrm{d}x.
\label{A2}
\end{multline}
Expanding the first integral in Eq.~(\ref{A2}), then applying integration by parts, and finally using Eq.~(\ref{eqn4}) gives 
\begin{align*}
-\frac{1}{2}\int \left(\chi_{+}^{\prime }+\chi_{-}^{\prime
}\right)^{2} \,\mathrm{d}x &=-\int {\chi_{+}^{\prime }}^{2}\,\mathrm{d}x-\int \chi_{+}^{\prime
}\chi_{-}^{\prime }\,\mathrm{d}x\\&=
\int \chi_{+}^{\prime\prime }\chi_{+}\,\mathrm{d}x-\int \chi_{+}^{\prime
}\chi_{-}^{\prime }\,\mathrm{d}x\\
&=\int \left[-\omega_{\epsilon} ^{2}+V^{\prime \prime }(u_{+})\right]\chi_{+}^{2}\,\mathrm{d}x-\int \chi_{+}^{\prime
}\chi_{-}^{\prime }\,\mathrm{d}x\\
&=-\omega_{\epsilon} ^{2}+\int V^{\prime \prime }(u_{+})\chi_{+}^{2}\,\mathrm{d}x-\int \chi_{+}^{\prime
}\chi_{-}^{\prime }\,\mathrm{d}x.
\end{align*}
Expanding the second integral in Eq.~(\ref{A2}) gives
\begin{multline*}
 -\int \frac{1}{2}V^{\prime \prime }\left(u_{+}+u_{-}-\frac{1}{2}\big[1+\tanh(qX)\big]\right)(\chi_{+}+\chi_{-})^{2}\,\mathrm{d}x\\
 =-\int V^{\prime \prime }\left(u_{+}+u
_{-}-\frac{1}{2}\big[1+\tanh(qX)\big]\right)\left(\chi_{+}^{2}+\chi_{+}\chi_{-}\right)\,\mathrm{d}x.
\end{multline*}
Adding the two integrals and rearranging the terms, Eq.~(\ref{A2}) becomes
\begin{multline*}
 a_2(X) = -\omega_{\epsilon} ^{2}+\int \left\{V^{\prime \prime }(u_{+})-V^{\prime \prime }\left(u_{+}+u_{-}-\frac{1}{2}\big[1+\tanh(qX)\big]\right)\right\}\chi_{+}^{2}\,\mathrm{d}x\\
-\int \chi_{+}^{\prime}\chi_{-}^{\prime}-V^{\prime \prime }\left(u_{+}+u
_{-}-\frac{1}{2}\big[1+\tanh(qX)\big]\right)\chi_{+}\chi_{-}\,\mathrm{d}x.
\end{multline*}
Note that for the reduced system, we only take the first term: $-\omega_{\epsilon}^{2}$ as the coefficient of $A^2$.

Next,
\begin{align*}
a_3(X) &=\int \frac{1}{2}\left[ u_{+}^{\prime }
-u_{-}^{\prime }-\frac{q}{2} \sech^2(qX)\right]^{2}\mathrm{d}x\\
 &=\int [u_{0}^{\prime }(x)]^{2}\,\mathrm{d}x-\int u_{+}^{\prime }u
_{-}^{\prime }\,\mathrm{d}x-\frac{q}{2}\int \left( u_{+}^{\prime }-u
_{-}^{\prime }\right) \sech^2(qX) \,\mathrm{d}x\\
 &\phantom{=}+\frac{q^2}{8}\int \sech^4(qX) \,\mathrm{d}x.
\end{align*}
Note that $a_3(X)=M_0+I(X)$, where $M_0$ and $I(X)$ are given in Eqs.~(\ref{eq:M0}) and (\ref{eq:I}), respectively.

Next,
\begin{equation*}
 a_4(X)=\int \frac{1}{2}(\chi_{+}+\chi_{-})^{2}\,\mathrm{d}x = \int {\chi_{+}}^{2}\, \mathrm{d}x + \int
\chi_{+}\chi_{-}\,\mathrm{d}x = 1 + \int \chi_{+}\chi_{-}\,\mathrm{d}x.
\end{equation*}
Note that for the reduced system, we only take the first term: $1$ as the coefficient of $\dot{A}^2$.

The remaining coefficients are
\begin{align*}
a_5(X)&=\int \left[ u_{+}^{\prime}-u_{-}^{\prime}-\frac{q}{2} \sech^2(qX)\right] \left(\chi_{+}+\chi_{-}\right)\,\mathrm{d}x\\
&=-2\int u_{-}^{\prime }\chi_{+}\,\mathrm{d}x -\int \frac{q}{2} \sech^2(qX)(\chi_{+}+\chi_{-})\,\mathrm{d}x,\\
a_6(X)&=\int \left(\chi_{+}+\chi_{-}\right)\left(\chi'_{+}-\chi'_{-}\right)\,\mathrm{d}x,\\
a_7(X)&=\int \left[ u_{+}^{\prime}-u_{-}^{\prime}-\frac{q}{2} \sech^2(qX)\right] \left(\chi'_{+}-\chi'_{-}\right)\,\mathrm{d}x,\\
a_8(X)&=\frac{1}{2}\int \left(\chi'_{+}-\chi'_{-}\right)^2\,\mathrm{d}x,\\
 a_9(X)&=-\frac{1}{6}\int V^{\prime \prime \prime}\left(u_{+}+u_{-}-\frac{1}{2}\big[1+\tanh(qX)\big]\right)(\chi_{+}+\chi_{-})^{3}\,\mathrm{d}x,\\
 a_{10}(X)&=- \frac{1}{24}\int V^{(iv) }\left(u_{+}+u_{-}-\frac{1}{2}\big[1+\tanh(qX)\big]\right)(\chi_{+}+\chi_{-})^{4}\,\mathrm{d}x,\\
 a_{11}(X)&=- \frac{1}{120}\int V^{(v) }\left(u_{+}+u_{-}-\frac{1}{2}\big[1+\tanh(qX)\big]\right)(\chi_{+}+\chi_{-})^{5}\,\mathrm{d}x,\\
 a_{12}(X)&=- \frac{1}{720}\int V^{(vi) }\left(u_{+}+u_{-}-\frac{1}{2}\big[1+\tanh(qX)\big]\right)(\chi_{+}+\chi_{-})^{6}\,\mathrm{d}x.
\end{align*}

Finally, the Euler--Lagrange equations (\ref{EL-basic}) for the CC Lagrangian~(\ref{nonreduced}) are
\begin{subequations}
\begin{equation}
\begin{aligned}
\big[a_5(X)+a_6(X)A\big]\ddot{X}+2a_4(X)\ddot{A} &= -2a'_4(X)\dot{X}\dot{A} + \big[a'_7(X)-a'_5(X)\big]\dot{X}^2\\
&\phantom{=} + \big[2a_8(X)-a'_6(X)\big]A\dot{X}^2 + a_1(X)+2a_2(X)A\\
&\phantom{=} + 3a_9(X)A^2 + 4a_{10}(X)A^3+5a_{11}(X)A^4+6a_{12}(X)A^5,
\end{aligned}
\end{equation}
\begin{equation}
\begin{aligned}
\big[2a_3(X)+2a_7(X)A+2a_8(X)A^2\big]\ddot{X}\\
 + \big[a_5(X)+a_6(X)A\big]\ddot{A} &= -a'_3(X)\dot{X}^2-a'_7(X)A\dot{X}^2 + \big[a'_4(X)-a_6(X)\big]\dot{A}^2\\
&\phantom{=} - 2a_7(X)\dot{A}\dot{X}-a'_8(X)A^2\dot{X}^2 - 4a_8(X)A\dot{A}\dot{X}\\
&\phantom{=} + a'_0(X)+a'_1(X)A +a'_2(X)A^2 + a'_9(X)A^3\\
&\phantom{=} + a'_{10}(X)A^4 + a'_{11}(X)A^5+a'_{12}(X)A^6.
\end{aligned}
\end{equation}
\label{ode-full}\end{subequations}
Here, primes denote derivatives with respect to the argument of the function, specifically $X$.
However, for the reasons discussed in \cite{weigel,weigel2}, we work with the reduced CC Lagrangian (\ref{lag1}), which corresponds to the system with the coefficients $\{a_i\}_{i=1,2,\hdots,5}$ in the main text above.

\acknowledgments
During the initial stages of this work, I.C.C.\ was partially supported by the LANL/LDRD Program through a Feynman Distinguished Fellowship at Los Alamos National Laboratory (LANL). LANL is operated by Los Alamos National Security, L.L.C.\ for the National Nuclear Security Administration of the U.S.\ Department of Energy under Contract No.\ DE-AC52-06NA25396. P.G.K.\ gratefully acknowledges support from the Alexander von Humboldt
Foundation, the Stavros Niarchos Foundation (via the Greek Diaspora Fellowship
Program) and the US National Science Foundation via grant PHY-1602994.

\end{document}